\documentclass[prb,showpacs,twocolumn,aps,superscriptaddress,a4paper]{revtex4-1}

\usepackage{color}
\usepackage{dcolumn}
\usepackage{amsmath}
\usepackage{graphicx}
\usepackage{latexsym}
\usepackage{amsfonts}
\usepackage{amssymb}
\DeclareGraphicsExtensions{.pdf,.gif,.jpg}

\newcommand{\be}{\begin{equation}}  
\newcommand{\ee}{\end{equation}}
\newcommand{\beq}{\begin{eqnarray}}  
\newcommand{\eeq}{\end{eqnarray}}  
\newcommand{\ket}[1]{| #1 \rangle}
\newcommand{\bra}[1]{\langle #1 |}
\newcommand{\ev}[2]{\langle #1 | #2 \rangle}
\newcommand{\mel}[3]{\langle #1 | #2 | #3 \rangle}

\renewcommand\Im{\operatorname{Im}}
\renewcommand\Re{\operatorname{Re}}

\def\ud{{\mathrm{d}}}
\def\a{{\alpha}}
\def\b{{\beta}}
\def\eps{{\epsilon}}
\def\g{\gamma}
\def\w{{\omega}}

\def\im{{\mathrm{i}}}
\def\ex{{\mathrm{e}}}
\def\intw{{\int\hspace{-2pt}\frac{\ud \omega}{2\pi}}}
\def\intwf{{\int\hspace{-2pt}\frac{\ud \omega}{2\pi}}f(\omega - \mu)}
\def\bG{\mbox{\boldmath $G$}}  
\def\bH{\mbox{\boldmath $h$}}  
\def\unit{\mbox{\boldmath $1$}}
\def\itS{\mbox{$\mathit{\Sigma}$}}
\def\itG{\mbox{$\mathit{\Gamma}$}}
\def\itL{\mbox{$\mathit{\Lambda}$}}
\def\itO{\mbox{$\mathit{\Omega}$}}
\def\itP{\mbox{$\mathit{\Pi}$}}

\begin{document}


\title{Time-dependent Landauer--B{\"u}ttiker formula: application to 
       transient dynamics in graphene nanoribbons}

\author{Riku Tuovinen}
\affiliation{Department of Physics, Nanoscience Center, FIN 40014, 
             University of Jyv{\"a}skyl{\"a}, Finland}

\author{Enrico Perfetto}
\affiliation{Dipartimento di Fisica, Universit\`a di Roma Tor Vergata, 
             Via della Ricerca Scientifica 1, I-00133 Rome, Italy}

\author{Gianluca Stefanucci}
\affiliation{Dipartimento di Fisica, Universit\`a di Roma Tor Vergata, 
             Via della Ricerca Scientifica 1, I-00133 Rome, Italy}
\affiliation{Laboratori Nazionali di Frascati, Istituto Nazionale di Fisica Nucleare, 
             Via E. Fermi 40, 00044 Frascati, Italy}
\affiliation{European Theoretical Spectroscopy Facility (ETSF)}

\author{Robert van Leeuwen}
\affiliation{Department of Physics, Nanoscience Center, FIN 40014, 
             University of Jyv{\"a}skyl{\"a}, Finland}
\affiliation{European Theoretical Spectroscopy Facility (ETSF)}

\date{\today}  

\begin{abstract}
In this work we develop a time-dependent extension of the Landauer--B\"uttiker approach to study 
transient dynamics in time-dependent quantum transport through molecular junctions. A key feature 
of the approach is that it provides a closed integral expression for the time-dependence of the 
density matrix of the molecular junction after switch-on of a bias in the leads or a perturbation
in the junction, which in turn, can be evaluated without the necessity of propagating 
individual single-particle orbitals or Green's functions. This allows for the study of 
time-dependent transport in large molecular systems coupled to wide band leads. As an 
application of the formalism we study the transient dynamics of zigzag and armchair graphene
nanoribbons of different symmetries. We find that the transient times can exceed several 
hundreds of femtoseconds while displaying a long time oscillatory motion
related to multiple reflections of the density wave in the nanoribbons at the ribbon--lead 
interface. This temporal profile has a shape that scales with the length of the ribbons and is
modulated by fast oscillations described by intra-ribbon and ribbon--lead transitions. Especially
in the armchair nanoribbons there exists a sequence of quasi-stationary states related to 
reflections at the edge state located at the ribbon--lead interface. In the case of zigzag 
nanoribbons there is a predominant oscillation frequency associated with virtual transitions 
between the edge states and the Fermi levels of the electrode. We further study the local bond
currents in the nanoribbons and find that the parity of the edges strongly affects the path of 
the electrons in the nanoribbons. We finally study the behavior of the transients for various 
added potential profiles in the nanoribbons. 
\end{abstract}

  
\maketitle  


\section{Introduction}\label{sec:intro}
The Landauer--B\"uttiker (LB) formalism\cite{landauer,buttiker} has been a real milestone in the 
quantum theory of charge transport. Its success is attributable to the simplicity of the LB 
equations, which provide a transparent and physically intuitive picture of the steady-state 
current, as well as to the possibility of combining the formalism with density-functional theory 
(DFT) for first principle calculations.
\cite{L.1995,LA.1998,TGW-1.2001,TGW-2.2001,BMOTS.2002,SA.2004,FWK.2004}  
Nevertheless, due to the raising interest in the microscopic understanding of ultrafast charge 
transfer mechanisms, the last decade has seen heightened effort in going beyond the (steady-state) 
LB formalism thus accessing the transient regime. Different time-dependent (TD) approaches have been
proposed to deal with different systems. Approaches based on the real-time propagation of scattering
states,\cite{KSARG.2005,QLLY.2006,SPC.2008,ZC.2009,SPC.2010,GWSHGW.2014} wave packets,
\cite{BCG.2008,CFPS.2009,B.2011} extended states with sharp boundaries
\cite{BSdV.2005,CEvV.2006,SBHdV.2007,EvV.2009} or complex absorbing potentials,
\cite{BSIN.2004,V.2011,ZCW.2013} and noninteracting Green's functions
\cite{MWG.2006,LY.2007,MGM.2007,PD.2008,CS.2009,ZCMKTYY.2010,XWW.2010,XJTetal.2012,ZXW.2012,PSS.2013,ZCC.2013,mceniry} 
are suited to include the electron--electron interaction in a DFT framework. Interactions can 
alternatively be treated using nonequilibrium diagrammatic perturbation theory and solving the 
Kadanoff--Baym equations for open systems.\cite{mssvl.2008,petriprb,friesen2} Several nonperturbative
methods have been put forward too but, at present, they are difficult to use for first-principle 
calculations. These include master-equation type approaches,
\cite{PW.2005,WSK.2006,WT.2009,MGM.2009,ZSKEB.2009,PW.2010,WZJY.2013} real-time path-integral 
methods,\cite{WETE.2008,SF.2009,SMR.2010,CR.2011} nonequilibrium renormalization group methods,
\cite{AS.2005,AS.2006,PSS.2010,WK.2010,APSSB.2011,KJKM.2012,ESGA.2012} the quantum-trajectory 
approach,\cite{O.2007,ASO.2009} the TD density matrix renormalization group
\cite{V.2004,S.2004,WF.2004,BSS.2010,S.2011} and the nonequilibrium dynamical mean field theory.
\cite{WOM.2009,WOEM.2009} 

In its original formulation the LB formalism treats the electrons as noninteracting. Indubitably, 
the neglection of the electron--electron and electron--phonon interactions is in many cases a too crude 
approximation. However, in the ballistic regime interaction effects play a minor role and the LB 
formalism is, still today, very useful to explain and fit several experimental curves. For instance 
the identification of the different transport mechanisms, the temperature dependence of the current,
the exponential decay of the conductance as a function of the length of the junction, etc. can all 
be interpreted within the LB formalism.\cite{cuevassheerbook} The TD approaches previously mentioned 
have the merit of extending the quantum transport theory to the time domain. However, they all are 
computationally more expensive and less transparent than the LB formalism 
{\em even for noninteracting electrons}. Therefore, considering the widespread use of the LB formalism 
in both the theoretical and experimental communities, it is natural to look for a TD-LB formula 
which could give the current at time $t$ at the same computational cost as at the steady state. 

For a single level initially isolated and then contacted to source and drain electrodes a TD-LB 
formula was derived by Jauho {\em et al.} in 1994.\cite{jauho} The treatment of the contacts in the 
initial state introduces some complications which, however, were overcame about ten years later.
\cite{stefanucci-rts} The approach of Ref.~\onlinecite{stefanucci-rts} was then applied to 
generalize the TD-LB formula to a single level with spin.\cite{perfetto} Nevertheless, only recently
we have been able to derive a TD-LB formula for arbitrary scattering regions.\cite{svlbook,rikuproc}
The only restriction of this formula is that the density of states of the source and drain 
electrodes is smooth and wide enough that the wide-band limit approximation (WBLA) applies. In this 
case one can derive a TD-LB formula not only for the total current but for the full one-particle 
density matrix. The explicit analytic result allows for interpretion of typical transient 
oscillations in terms of electronic transitions within the molecular junction or between the 
junction and the leads, as well as the different damping times. Owing to the low computational cost 
one can consider very large systems and arbitrarily long propagation times. 

In this work we briefly review the results of Refs.~\onlinecite{svlbook,rikuproc} and generalize 
them to include arbitrary perturbations in the molecular junction. 
We further present a convenient implementation scheme to extract densities and local currents, and 
demonstrate the feasibility of the method in graphene nanoribbons (GNR).
\cite{CGPNG.2009,katsnelsonbook,onipko,1468-6996-11-5-054504} So far, real-time investigations of 
GNRs have been limited to small size\cite{XKZJZYC.2013} and weak biases.\cite{PSC.2010} As the TD-LB 
formalism is not limited to weak driving fields we could study the transient dynamics in the unexplored 
strong bias regime. In GNRs there are plenty of interesting nanoscale size effects depending on the 
topology of the edges. Our main findings are that for large biases (i) the time to relax to the steady 
state exceeds hundreds of femtoseconds; (ii) in the transient current and density of zigzag GNRs there 
is a predominant oscillation frequency associated with virtual transitions between the edge states and 
the Fermi levels of the electrodes; (iii) the currents in the armchair GNRs exhibit a sequence of 
quasi-stationary states whose duration increases with the length of the GNR; and (iv) the parity of 
the edges strongly affects the path of the electrons inside the GNR.

The paper is organized as follows. In Section \ref{sec:theory} we introduce the system and present 
the main results of the TD-LB formalism. Here we also illustrate the implementation scheme and defer
the numerical details to the Appendix. The TD results on GNRs are collected in Section 
\ref{sec:results} where we investigate the effects of the edge states, the quasi-stationary 
currents, the even--odd parity effect on the current--density profile and a perturbed GNR. 
Finally we draw our conclusions in Section \ref{sec:conclusion}.
 

\section{Theoretical background}\label{sec:theory}
\subsection{System set-up and earlier work}
\label{set-upsec}
\begin{figure}
\includegraphics[width=0.3\textwidth]{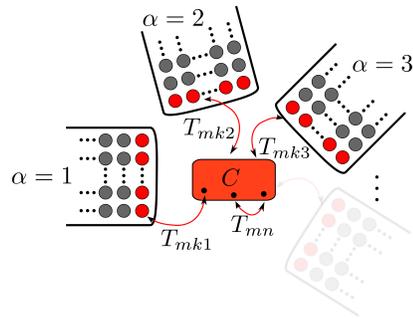}
\caption{Schematic of the quantum transport setup: a noninteracting central region, $C$, is 
         coupled to an arbitrary number of leads.}
\label{fig:schematic}
\end{figure}

We investigate quantum transport between metallic wide-band leads
and a noninteracting central region. The setup is otherwise
as general as possible; the number and the structure of the leads are
arbitrary as is the size and the structure of the central region. 
The Hamiltonian is of the form
\beq\label{eq:nonintham}
\hat{H} & = &
\sum_{k\a,\sigma}\eps_{k\a}  \hat{d}_{k\a,\sigma}^\dagger \hat{d}_{k\a,\sigma}
+
\sum_{mn,\sigma}T_{mn} \hat{d}_{m,\sigma}^\dagger\hat{d}_{n,\sigma} \nonumber \\
& + & 
\sum_{mk\a,\sigma}\left(T_{mk\a} \hat{d}_{m,\sigma}^\dagger\hat{d}_{k\a,\sigma}
                           +T_{k\a m} \hat{d}_{k\a,\sigma}^\dagger\hat{d}_{m,\sigma}\right)
\ .
\eeq
Here $\sigma$ is a spin index and $k\a$ denotes the $k$th
basis function of the $\a$th lead while $m$ and $n$ label basis states
in the central region. The corresponding creation and annihilation operators for these
states are denoted by $\hat{d}^\dagger$ and $\hat{d}$, respectively. The single-particle levels of
the leads are given by $\eps_{k\a}$ while the matrices $T$ give the hoppings between
the molecular and molecule--lead states. 
This is depicted schematically in Fig.~\ref{fig:schematic}.

At times $t<t_0$ the system is in thermal equilibrium at inverse
temperature $\b$ and chemical potential $\mu$, the density matrix
having the form $\hat{\rho} = \frac{1}{\mathcal{Z}}\ex^{-\b(\hat{H} -\mu\hat{N})}$ where
$\mathcal{Z}$ is the grand-canonical partition function of the connected lead--molecule system. 
At $t=t_0$ a sudden bias of the form
\be
\hat{V} = \theta (t-t_0) \sum_{k\a,\sigma } V_\a 
\hat{d}_{k\a,\sigma}^\dagger \hat{d}_{k\a,\sigma} \nonumber
\ee
is applied to leads, where $V_\a$ is the bias strength in lead $\a$. This potential
drives the system out of equilibrium and charge carriers start to 
flow through the central region. To calculate the time-dependent 
current we use the equations of motion for the one-particle Green's 
function on the Keldysh contour $\gamma_{\text{K}}$. This quantity is 
defined as the ensemble average of the contour-ordered product of 
particle creation and annihilation operators in the Heisenberg 
picture \cite{svlbook}
\be\label{eq:greenf}
G_{rs}(z,z') =
-\im\langle\mathcal{T}_{\g_{\mathrm{K}}}[\hat{d}_{r,\mathrm{H}}(z)\hat{d}_{s,\mathrm{H}}^\dagger(z')]\rangle
\ee
where the indices $r$, $s$ can be either indices in the leads or in 
the central region and the
variables $z$, $z'$ run on the contour. This
contour has a forward and a backward branch on the real-time axis,
$[t_0,\infty[$, and also a vertical branch on the imaginary axis,
$[t_0,t_0-\im\b]$  describing the
initial preparation of the system \cite{petriprb}. 
The matrix $\bG$ with matrix elements $G_{rs}$ satisfies the 
equations of motion \cite{kb}
\beq
\left[\im\frac{\ud}{\ud z} - \bH(z)\right]\bG(z,z') & = & \delta(z,z')\unit
\label{eq:eom-left} \ , \\
\bG(z,z')\left[-\im \frac{\stackrel{\leftarrow}{\ud}}{\ud z'} - \bH(z')\right] & = &
\delta(z,z')\unit \label{eq:eom-right} \ ,
\eeq
with Kubo--Martin--Schwinger boundary conditions, i.e. the Green's function 
is anti-periodic along the contour. Here  
$\bH(z)$ is the single-particle Hamiltonian.
In the basis $k\a$ and $m$ the matrix $\bH$ has
the following block structure
\be\label{eq:ham-matrix}
\bH = \begin{pmatrix}h_{1 1} & 0 & 0 & \cdots & h_{1 C} \\
                   0 & h_{2 2} & 0 & \cdots & h_{2 C} \\
                   0 & 0 & h_{3 3} & \cdots & h_{3 C} \\
                   \vdots & \vdots & \vdots & \ddots & \vdots \\
                   h_{C 1} & h_{C 2} & h_{C 3} & \cdots &
h_{C C}
    \end{pmatrix} \ ,
\ee
where $(h_{\a \a'})_{kk'} = \delta_{\a\a'}\delta_{kk'}\eps_{k\a}$
corresponds to the leads, $(h_{\a C})_{km} = T_{k\a m}$ is the
coupling part, and $(h_{C C})_{mn} = T_{mn}$ accounts for the
central region. We approximate the retarded embedding self-energy 
as a purely imaginary constant, according to WBLA
\beq
\label{eq:wbla}
\itS^{\textrm{R}}_{\a,mn}(\w) & = & \sum_k T_{m k\a} \, \frac{1}{\w-\eps_{k\a}-V_\a+ \im\eta} \, T_{k\a n} \nonumber \\
& = & -\frac{\im}{2}\itG_{\a,mn} \ .
\eeq
In other words, the level-width functions $\itG_\a$ appear as the wide-band approximation 
for the retarded embedding self-energy $\itS_\a^{\text{R}}(\w) = -\im\itG_\a/2$ for 
which $\itG = \sum_\a \itG_\a$. Due to the coupling between the central region and the 
leads the matrix $\bG$ has nonvanishing entries everywhere
\be\label{eq:green-self-matrix}
\bG = \begin{pmatrix} G_{11} & \cdots & G_{1 C} \\
                    \vdots & \ddots & \vdots \\
                    G_{C 1} & \cdots & G_{C C}
    \end{pmatrix} .
\ee
The equations of motion \eqref{eq:eom-left} and~\eqref{eq:eom-right}
for the Green's function $G_{CC}$ projected onto the central region have
been solved analytically in WBLA\cite{rikuproc} to give the time-dependent one-particle
reduced density matrix (TD1RDM) as the 
equal-time limit $\rho(t) = -\im G_{CC}^<(t,t)$:
\beq\label{eq:td-dens}
\rho(t) & = & \intwf \sum_{\a} \Big\{ \ A_{\a}(\w +
V_{\a})\nonumber \\
& + & V_{\a}\left[\ex^{\im(\w + V_{\a}-h_{\mathrm{eff}})t} G^{\mathrm{R}}(\w)A_\a(\w + V_{\a}) +
\mathrm{h.c.}\right] \nonumber \\
& + & V_{\a}^2\ex^{-\im h_{\mathrm{eff}} t}G^{\mathrm{R}}(\w) A_\a(\w +
V_{\a})G^{\mathrm{A}}(\w)\ex^{\im h_{\mathrm{eff}}^\dagger t} \ \Big\} 
\eeq
where $f$ is the Fermi function, $G^{\mathrm{R}}(\w)=(\w-h_{\mathrm{eff}})^{-1}$ and 
$G^{\mathrm{A}}(\w) = [G^{\mathrm{R}}(\w)]^\dagger$ are the retarded and advanced Green's 
functions, $h_{\mathrm{eff}} = h_{CC} - \im\itG/2$ is the effective 
single-particle Hamiltonian, and the partial spectral functions are 
$A_\a(\w) =  G^{\mathrm{R}}(\w)\itG_{\a}G^{\mathrm{A}}(\w)$. The full spectral function is then simply
$A(\w)=\sum_{\a}A_{\a}(\w)$.

We emphasize that Eq. \eqref{eq:td-dens} is an explicit closed formula for the 
equal-time $G^{<}$ or, equivalently, for the TD1RDM. All the quantities 
inside the integral can be calculated  without 
the need of storing auxiliary quantities at earlier times.
In other words, if we want to know the TD1RDM at time $t$ we simply 
need to evaluate the integral in Eq. (\ref{eq:td-dens}). As no 
propagation is required we have access to nonequilibrium 
quantities at arbitrary times after the switch-on of the bias. 
This is the most important feature of Eq. \eqref{eq:td-dens}. In 
fact, for large and weakly coupled junctions the transient regime can 
exceed several hundreds of femtoseconds and, at present, these 
time-scales are out-of-reach of the available TD approaches. 

Similarly to the TD1RDM the time-dependent current through the interface of the 
$\a$th lead has an explicit closed expression which generalizes the 
LB formula to the time domain\cite{rikuproc}
\begin{widetext}
\beq\label{eq:td-current}
I_\a(t) & = & -2 \intwf \sum_{\b} \mathrm{Tr} \Big\{ \itG_{\a} G^{\mathrm{R}}(\w + V_{\b}) \itG_{\b} G^{\mathrm{A}}(\w + V_{\b}) - \itG_{\a} G^{\mathrm{R}}(\w + V_{\a})\itG_{\b} G^{\mathrm{A}}(\w + V_{\a}) \nonumber \\
& + & V_{\b}\left[\itG_{\a}\ex^{\im(\w + V_{\b} -
h_{\mathrm{eff}})t}G^{\mathrm{R}}(\w)\left(-\im\delta_{\a\b}G^{\mathrm{R}}(\w + V_{\b}) + A_\b(\w +
V_{\b})\right) + \mathrm{h.c.} \right]\nonumber \\
& + & V_{\b}^2\itG_{\a}\ex^{-\im h_{\mathrm{eff}} t}G^{\mathrm{R}}(\w)A_\b(\w +
V_{\b})G^{\mathrm{A}}(\w)\ex^{\im h_{\mathrm{eff}}^\dagger t}\ \Big\}
\eeq
\end{widetext}
where $\b$ runs over all the leads. As $h_{\mathrm{eff}}$ 
is non-hermitian the terms in the last two rows of Eq. 
(\ref{eq:td-current}) vanish exponentially when $t\to\infty$ and one 
recovers the steady-state LB formula. It is easy to verify that the 
current correctly vanishes for all $t$ at zero bias, $V_{\a}=0$, and 
for $t=0$ at any bias. In the remainder of this Section we present 
a convenient numerical procedure to evaluate Eq. (\ref{eq:td-dens}) 
as well as a generalization of the same formula to include arbitrary 
spatially-dependent perturbations in the central region.

\subsection{Expansion in the $h_{\mathrm{eff}}$ eigenbasis}\label{sec:expansion}
We expand the result in Eq.~\eqref{eq:td-dens} in the eigenbasis of the
non-hermitian effective Hamiltonian $h_{\mathrm{eff}}$. This object has separate left and
right eigenvectors forming a mutually \emph{biorthogonal} set
$\{  \ket{\Psi_j^{\mathrm{L}}} , \ket{\Psi_j^{\mathrm{R}}}\}$ with
\beq\label{eq:eval-eq}
\begin{cases}
\bra{\Psi_j^{\mathrm{L}}}h_{\mathrm{eff}} = \eps_j\bra{\Psi_j^{\mathrm{L}}} \\
h_{\mathrm{eff}}\ket{\Psi_j^{\mathrm{R}}} = \eps_j\ket{\Psi_j^{\mathrm{R}}} \ .
\end{cases}
\eeq
By the biorthogonality we have $\ev{\Psi_j^{\text{L}}}{\Psi_k^{\text{R}}} = \delta_{jk} \ev{\Psi_j^{\text{L}}}{\Psi_j^{\text{R}}}$,
where we can choose an appropriate normalization of the diagonal elements.

We notice that in Eq.~\eqref{eq:td-dens} in every term there is $h_{\mathrm{eff}}$ on the left and
$ h_{\mathrm{eff}}^\dagger$ on the right. This in mind, and looking at how the matrix operates in 
Eq.~\eqref{eq:eval-eq} we choose to expand in the `left--left' 
eigenbasis, i.e., we multiply
the density matrix in Eq.~\eqref{eq:td-dens} from left with a row vector 
$\bra{\Psi^{\text{L}}}$ and from the right by a column vector $\ket{\Psi^{\text{L}}}$.
In order to calculate a matrix element 
$\mel{m}{\rho(t)}{n}$ in the original basis of region $C$
we insert a complete set of left and right eigenvectors 
of $h_{\mathrm{eff}}$. 
The resolution of identity reads
\be
\unit = \sum_j \frac{\ket{\Psi_j^{\text{R}}} \bra{\Psi_j^{\text{L}}}}{\ev{\Psi_j^{\text{L}}}{\Psi_j^{\text{R}}}} = \sum_j \frac{\ket{\Psi_j^{\text{L}}} \bra{\Psi_j^{\text{R}}}}{\ev{\Psi_j^{\text{R}}}{\Psi_j^{\text{L}}}} 
\ee
and hence
\be\label{eq:solution}
\mel{m}{\rho(t)}{n} = \sum_{j,k} \frac{\ev{m}{\Psi_j^{\text{R}}}}{\ev{\Psi_j^{\text{L}}}{\Psi_j^{\text{R}}}}
             \frac{\ev{\Psi_j^{\text{R}}}{n}}{\ev{\Psi_k^{\text{R}}}{\Psi_k^{\text{L}}}}
             \mel{\Psi_j^{\text{L}}}{\rho(t)}{\Psi_k^{\text{L}}} \ .
\ee
The matrix elements 
$\rho_{jk}(t)=\mel{\Psi_j^{\text{L}}}{\rho(t)}{\Psi_k^{\text{L}}}$
can easily be extracted from Eq. (\ref{eq:td-dens}) and read
\beq\label{eq:td-dens-decomposed}
\rho_{jk}(t) & = & \sum_\a\itG_{\a,jk}\itL_{\a,jk} \nonumber \\
& + & \sum_\a V_{\a}\itG_{\a,jk}\left[\itP_{\a,jk}(t)+\itP_{\a,kj}^*(t)\right] \nonumber \\
& + & \sum_\a V_{\a}^2\itG_{\a,jk}\ex^{-\im(\eps_j-\eps_k^*)t}\itO_{\a,jk}
\eeq
with
\be
\itG_{\a,jk} =  \mel{\Psi_j^{\text{L}}}{\itG_\a}{\Psi_k^{\text{L}}}
\ee
and
\be
\itL_{\a,jk} = \intw\frac{f(\omega - 
\mu)}{(\w+V_{\a}-\eps_j)(\w+V_{\a}-\eps_k^*)} \ ,
\label{eq:terms1}
\ee
\be
\itP_{\a,jk}(t) = \intw\frac{f(\omega - 
\mu)\ex^{\im(\w+V_{\a}-\eps_j)t}}{(\w-\eps_j)(\w+V_{\a}-\eps_j)(\w+V_{\a}-\eps_k^*)} \ ,
\label{eq:terms2}
\ee
\be
\itO_{\a,jk} =  \intw\frac{f(\omega - \mu)}{(\w-\eps_j)(\w+V_{\a}-\eps_j)(\w+V_{\a}-\eps_k^*)(\w-\eps_k^*)} \ .
\label{eq:terms3}
\ee

The first row of Eq.~\eqref{eq:td-dens-decomposed} gives the  
steady-state value of the TD1RDM. The time dependent part is 
contained in the functions $\itP$ in the second row 
and in the exponential in the third row. By inspection of Eq. 
(\ref{eq:td-dens-decomposed}) we 
see that
transitions between the leads and the central region are described by the terms $\itP$ 
(oscillations of frequency $\w_j = |V_{\a}-\Re \eps_j|$), whereas transitions within the central
region are described by the exponential term in the third line (oscillations of frequency 
$\w_{jk} = |\Re \eps_j-\Re \eps_k|$).\cite{rikuproc} As the 
eigenvalues $\eps_j$ are, in general, complex we infer that 
electronic transitions between states in the central region are 
damped faster than those involving states at the Fermi energies 
$\mu+V_{\a}$. 
In the zero-temperature limit the integrals in 
Eq.~(\ref{eq:terms1}-\ref{eq:terms3}) are given in terms of logarithms and
exponential integral functions (of complex variable), which can be evaluated using 
an extremely accurate numerical algorithm proposed recently in the 
context of computer graphics,\cite{expint} see Appendix \ref{app:formulae}.

\subsection{Switching on of electric and magnetic fields in the central region}
The TD1RDM of Eq. (\ref{eq:td-dens}) and the TD current of Eq. 
(\ref{eq:td-current}) refer to systems driven out of equilibrium by 
an external bias. Here we generalize these results to 
include the sudden switch-on of electric and/or magnetic fields in the 
central region. We consider the system described in Section 
\ref{set-upsec} with central-region Hamiltonian $h_{CC}$ in equilibrium 
and $\widetilde{h}_{CC}$ for $t>t_0$, where $t_{0}$ is the time at which 
the bias is switched on. The switch-on of an electric field is useful to study, e.g., the 
effects of a gate voltage or to model the self-consistent voltage 
profile within the central region. In this case 
\be
(\widetilde{h}_{CC})_{mn} = T_{mn} +  u_{mn}
\ee
where $u_{mn}$ are the matrix elements of the scalar potential 
between two basis states of the central region. The switch-on of 
a magnetic field is instead useful to study, e.g., the Aharonov--Bohm
effect in ring geometries or the Landau levels in planar 
junctions like graphene nanoribbons. In this case
\be
(\widetilde{h}_{CC})_{mn} = T_{mn}\ex^{\im\a_{mn}}
\ee
where the sum of the Peierls phases $\a_{mn}=-\a_{nm}$ 
along a closed loop yields the magnetic flux (normalized to the flux 
quantum $\phi_{0}=h/2e$) across the loop. 

Having two different Hamiltonians for the central region ($h_{CC}$ at 
times $t<t_{0}$ and $\widetilde{h}_{CC}$ at times $t>t_{0}$), we need to adjust the 
derivation worked out in the earlier study in 
Ref.~\onlinecite{rikuproc}. By definition the
Matsubara Green's function remains unchanged since it only depends on 
the Hamiltonian at times $t<t_{0}$. On the other hand, for Green's functions 
having components on the horizontal 
branches of the Keldysh contour,  we have to use the
Hamiltonian $\widetilde{h}_{CC}$. The calculations are rather lengthy but 
similar to those presented in Ref.~\onlinecite{rikuproc}; we outline the main steps in 
Appendix~\ref{app:gate} and state here only the final result for the TD1RDM
\beq\label{eq:td-dens-gate}
\rho(t) & = & \intwf\sum_\a\left\{ \widetilde{A}_\a(\w+V_{\a}) \right. \nonumber \\
& & \left. + \left[\ex^{\im(\w+V_{\a}-\widetilde{h}_{\mathrm{eff}})t}G^{\mathrm{R}}(\w)\widetilde{V}_{\a}\widetilde{A}_\a(\w+V_{\a}) + \mathrm{h.c.} \right]  \right. \nonumber \\
& & \left. +\ex^{-\im \widetilde{h}_{\mathrm{eff}} t}G^{\mathrm{R}}(\w)\widetilde{V}_{\a}\widetilde{A}_\a(\w+V_{\a})\widetilde{V}_{\a}^\dagger G^{\mathrm{A}}(\w)\ex^{\im \widetilde{h}_{\mathrm{eff}}^\dagger t} \right\} \ , \nonumber \\
\eeq
where the functions with a \emph{tilde} signify that they are  
calculated using $\widetilde{h}_{CC}$, except
for $\widetilde{V}_{\a} = V_{\a} \unit - (\widetilde{h}_{CC}-h_{CC})$
 which is to be understood as a matrix in this case 
(in Eq.~\eqref{eq:td-dens} it was proportional to the identity 
matrix). The retarded/advanced Green's functions in 
Eq.~\eqref{eq:td-dens-gate} do not have tilde since they originate 
from the analytic continuation of $G^{\mathrm{M}}$.
In the limit $\widetilde{h}_{CC} \to h_{CC}$ it 
is easy to check that the results in Eqs.~\eqref{eq:td-dens} and~\eqref{eq:td-dens-gate} agree.

For the case of perturbed central 
region we would also like to have a similar result  as in 
Eq.~\eqref{eq:td-dens-decomposed}. Since $h_{\rm eff}$ and 
$\widetilde{h}_{\rm eff}$ do not necessarily commute 
the left/right eigenstates are not the same. For instance, in the 
second row of Eq.~\eqref{eq:td-dens-gate} we need to insert a complete set of left/right 
eigenstates of $h_{\mathrm{eff}}$ (resolution of the identity) in between the first exponential and 
$G^{\mathrm{R}}$, and so on. This leads to extra sums and overlaps 
between different bases. The resulting generalization of Eq. 
(\ref{eq:td-dens-decomposed})  is derived in 
Appendix~\ref{app:gate}.

\subsection{Physical content of the TD1RDM}
From the TD1RDM in the left--left basis we can extract the matrix 
elements in the site basis according to Eq. (\ref{eq:solution}).
In the site basis the diagonal elements
give the site densities (or local occupations) of the central region. 
The off-diagonal elements are instead related to the bond currents 
and the kinetic energy density.\cite{SPC.2010,T.2011}
The site densities and the bond currents are related by the 
continuity equation
$\partial_t n_m = \sum_n I_{mn}$, stating that the currents flowing in and out of 
site $m$ must add up to the temporal change of density in that site. 
It is easy to show that the bond currents are given by
\be\label{eq:bondcurrent}
 I_{mn} = 2 \,  \Im \left[T_{mn}\ex^{\im\a_{mn}}\rho_{nm}\right] \ .
\ee
At the steady-state ($t\to\infty$) one can verify that our equations 
for the TD1RDM
correctly imply $\sum_n I_{mn}=0$.


\section{Results}\label{sec:results}
We implement the framework described in the previous Section and in 
the Appendices to study the transient dynamics of GNRs coupled to 
metallic leads in the zero-temperature limit.
We are especially interested to investigate the so far unexplored 
region of large biases, where the Dirac (low-energy) Hamiltonian 
is inadequate.  By looking at 
time-dependent quantities, such as densities and bond currents, we 
perform a sort of  \emph{spectroscopical} analysis by discrete 
Fourier transforming the transient curves and reveal the dominant transitions 
responsible for the slow relaxation to a steady state.

\begin{figure}
\includegraphics[width=0.45\textwidth]{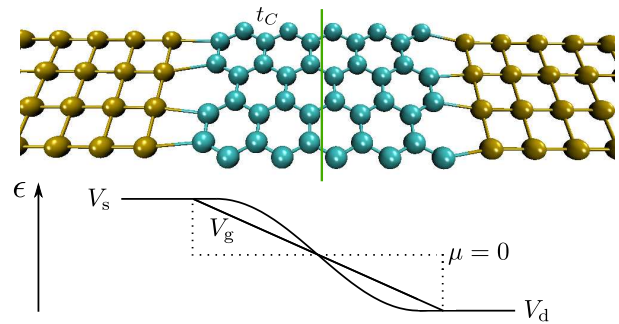}
\caption{Transport setup of a (zigzag) graphene nanoribbon connected to metallic leads: 
         contacts to leads are between doubly-coloured bonds; bridge (explained in text) 
         is shown by the green cutting line. The structure of the 
	 leads is shown for illustratory purposes. Voltage profile 
         is shown below the structure.}
\label{fig:setup}
\end{figure}

The transport setup is shown in Fig.~\ref{fig:setup}.
The leads are semi-infinite with terminal sites coupled to a 
GNR. The GNR is modeled by a single-orbital $\pi$-electron network, 
parametrized by nearest neighbor
hopping $t_{C}=-2.7$ eV;\cite{AriHarju} second and third 
nearest neighbour hoppings\cite{AriHarju} are neglected but can be included at the 
same computational price. The size and the orientation [zigzag (zGNR), armchair (aGNR)] 
of the GNR can be chosen freely as well as the structure of the leads. The strength of the 
level-width functions, $\itG_{\a}$, depends on both the couplings to the leads and the internal 
properties of the leads. Even though in our framework $\itG_{\a}$ can be 
any positive semidefinite matrix\cite{Schomerus,Blanter} here we take it of the form
\be
\itG_{\a,mn} = \gamma_\a  \, \Delta_{\a,mn} 
\ee
where $\Delta_{\a,mn}=\delta_{mn}$ when $m,n$ labels edge atoms 
contacted to
lead $\a$ and $\Delta_{mn,\a}=0$ otherwise. In our calculations we 
choose $\gamma_\a = 0{.}1$ eV independent of $\a$.
The chemical potential is set to $\mu=0$ in order to have a charge 
neutral GNR in equilibrium. The system is driven out of equilibrium by a sudden 
symmetric bias voltage between source and drain electrodes, i.e.,  
$V_{\a} = \pm V_{\text{sd}}/2$. 
The strength of the potential profile within the central region  is of amplitude 
$V_{\text{g}}$ and can be, e.g., linear or sinusoidal as illustrated 
in Fig.~\ref{fig:setup}, or of any other shape. 
To analyze the output of the numerical simulations 
we consider a cutting line or a {\em bridge} in the middle of the GNR and calculate 
the sum of all bond currents for the bonds cut by the bridge, see Fig.~\ref{fig:setup}.
In the following this sum of bond currents is denoted by $I$.
We measure energies in units of  $\eps = 1$ eV and therefore the unit
of time $t = \hbar/\eps \approx 6{.}58\cdot10^{-16}$ seconds and the unit of current
$I = e \eps/\hbar \approx  2{.}43\cdot10^{-4}$ amperes.

\subsection{Transient spectroscopy of zGNR and aGNR}\label{sec:transient}

\begin{figure}[t]
\includegraphics{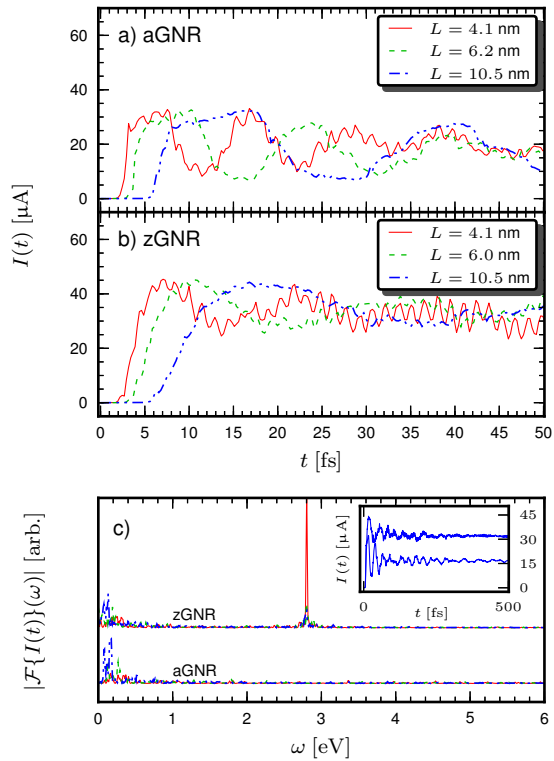}
\caption{Time-dependent bond currents through ribbons of \emph{varying length}:
         a) aGNR: (fixed width $W = 1{.}5$ nm (13)), 
         b) zGNR: (fixed width $W = 1{.}6$ nm (8)), and
         c) the corresponding Fourier transforms (zGNR is offset for clarity);
         the inset shows the long-time behaviour of the currents for $L=10{.}5$ nm in a) and b).
         [The line colours and styles correspond to those in a) and b).]}
\label{fig:length}
\end{figure}

\begin{figure}[t]
\includegraphics{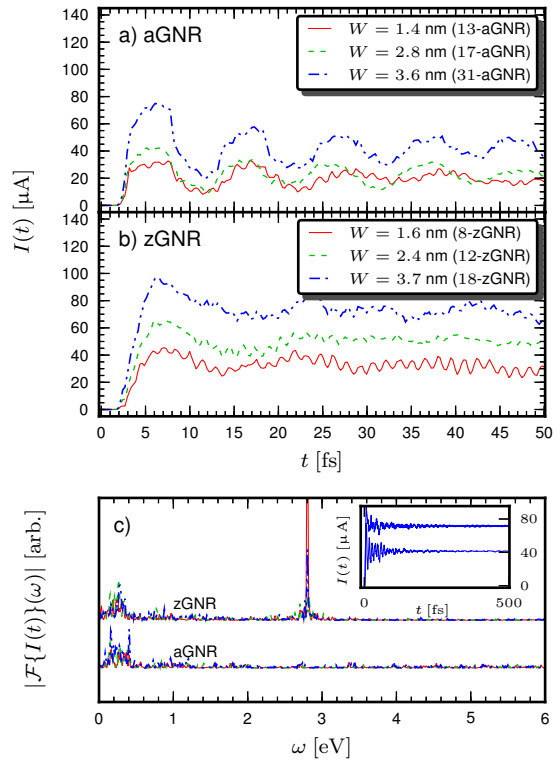}
\caption{Time-dependent bond currents through ribbons of \emph{varying width}:
         a) aGNR: (fixed length $L = 4{.}1$ nm), 
         b) zGNR: (fixed length $L = 4{.}1$ nm), and
         c) the corresponding Fourier transforms (zGNR is offset for clarity);
         the inset shows the long-time behaviour of the currents for $W = 3{.}6$ nm in a) 
         and $W = 3{.}7$ nm in b), respectively. 
         [The line colours and styles correspond to those in a) and b).]}
\label{fig:width}
\end{figure}

Let us study the dependence of the TD current on the length of the GNR at fixed 
width and bias voltage. 
For aGNRs of width $1{.}4$ nm (this is a $13$-aGNR where $13$ refers to 
the number of armchair dimer rows\cite{width}) and a zGNRs of width $1{.}6$ nm 
(this is an $8$-zGNR where $8$ is the number of zigzag rows\cite{width}) we show $I$ in 
Fig.~\ref{fig:length}a-b and the Fourier transforms in 
Fig.~\ref{fig:length}c. The Fourier transforms
are calculated from the long-time simulations shown in 
the inset of Fig.~\ref{fig:length}c where we subtract the steady-state value from the
sample points, take the absolute value of the 
result and use Blackman-window filtering.\cite{blackman} 
In both cases the bias voltage is $V_{\text{sd}} = 5{.}6$ eV  and 
$V_{\text{g}}=0$ eV.
By increasing the length of the ribbon the initial transient starts with 
a delay, since the current is measured in the center (see Fig.~\ref{fig:setup}), but the 
steady-state value is roughly the same.
The overall number of states also increases, and hence, more states close to the Fermi 
level are available as transport channels. Consequently smaller transition energies 
become dominant and the peaks in the Fourier spectra shift towards 
smaller frequencies.
For the zGNRs we also find a high-energy peak independent of the 
length; this peak is responsible for the fast superimposed oscillations in the time 
domain.   
The peak appears at frequency $\w = V_{\text{sd}}/2 = 2{.}8$ eV 
and therefore corresponds to transitions between the lead Fermi 
energy and zero-energy states in the ribbon, i.e., the edge states.
The edge states are  weakly coupled to the leads and therefore 
these transitions are slowly damped. 
As a matter of fact similar high-frequency oscillations are visible 
in aGNRs as well, see panel a. Nevertheless, the Fourier transform 
does not show any high frequency peak in this case. In aGNRs we have zigzag edges 
at the interface and hence edge states  strongly coupled to the 
leads. The high-frequency oscillations in aGNRs are damped 
faster than in zGNRs, see panel c, and are not visible in 
the Fourier spectrum. 

\begin{figure}[t]
\includegraphics{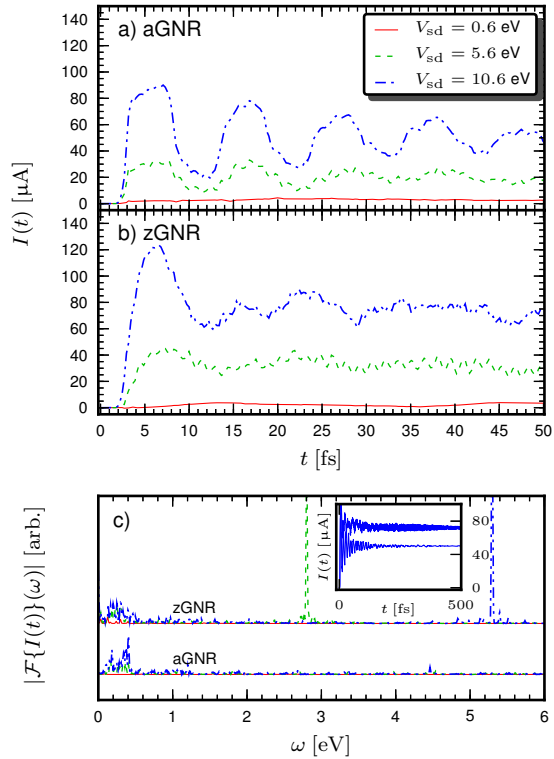}
\caption{Time-dependent bond currents through fixed-size ribbons with \emph{varying bias voltage}
         a) aGNR: ($W = 1{.}5$ nm (13), $L = 4{.}1$ nm), 
         b) zGNR: ($W = 4{.}1$ nm (8), $L = 4{.}1$ nm), and
         c) the corresponding Fourier transforms (zGNR is offset for clarity);
         the inset shows the long-time behaviour of the currents for $V_{\mathrm{sd}}=10{.}6$ 
         eV in a) and b).}
\label{fig:voltage}
\end{figure}

\begin{figure}[t]
\includegraphics[width=0.5\textwidth]{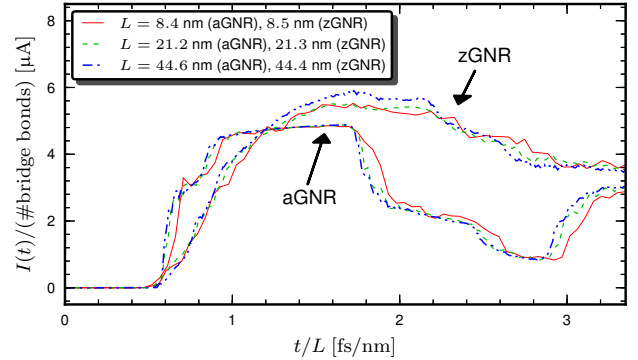}
\caption{First transients of the time-dependent current through ribbons of varying length
         divided by the number of bonds in the bridge. The horizontal axis is scaled by 
         the length of the corresponding ribbon.}
\label{fig:plateau}
\end{figure}

Next we vary the width of the ribbons while keeping the length and the bias voltage fixed. In 
Fig.~\ref{fig:width} we show the dependency
on the width for aGNRs and zGNRs of length 
$4{.}1$ nm. Depending on the width the ribbon is either metallic or 
semiconducting.\cite{1468-6996-11-5-054504} However, as the gap in 
the semiconducting case is much smaller than the applied voltage 
$V_{\text{sd}} = 5{.}6$ eV the conducting properties are not affected 
by the gap.
When increasing the width of the ribbon the length of the bridge, through which 
the cumulative bond current $I$ is calculated, increases and so does the 
steady-state value of $I$. However, the transient features remain the same 
as clearly illustrated in the Fourier spectrum of panel c. 
Thus, at difference with the results of Fig. \ref{fig:length}c,
the widening of the ribbon does not cause a shift of the low-energy peaks toward 
smaller energies. As expected, this is true also for the high-energy peak in 
zGNRs, in agreement with the fact that the energy of the edge-states 
is independent of the size of the ribbon. 

As a third case we study the effect of increasing the bias voltage 
(while still keeping $V_{\rm g}=0$). In Figs.~\ref{fig:voltage}a and~\ref{fig:voltage}c we 
show the results for 
$13$-aGNR of length $4{.}1$ nm and width $1{.}4$ nm, and in Figs.~\ref{fig:voltage}b 
and~\ref{fig:voltage}c the results for $8$-zGNR of length $4{.}1$ nm and width $1{.}6$ nm 
(ribbons of comparable sizes). For zGNR the frequency of 
the oscillations associated
to the edge-state transitions increases linearly with the bias, as it 
should be. We 
also observe that for 
both ribbons  the transient regime lasts longer the 
larger is the bias, and that the steady-state is attained after
several hundreds of femtoseconds. 

As a general remark of all the simulations shown in this subsection 
we can say that the absolute values of the 
steady-state currents are higher through zGNRs than through aGNRs (of comparable sizes). 
It is not easy to provide an intuitive explanation of this observation 
since at large biases there are very many states which contribute to the absolute value of the 
steady-state current. We also observe that
the micro--milliampere range for the current with bias in the eV range 
agrees with the experimental results of  Refs.~
\onlinecite{PhysRevLett.100.206803, 0957-4484-22-26-265201, Science.319.1229, 
doi:10.1021/nl2023756, doi:10.1021/nl070133j, doi:10.1109/TED.2007.891872, 
doi:10.1038/nnano.2008.268, PhysRevLett.100.206802}. 

\subsection{Quasi-stationary currents}\label{sec:plateau}

\begin{figure*}[ht!]
\includegraphics[width=\textwidth]{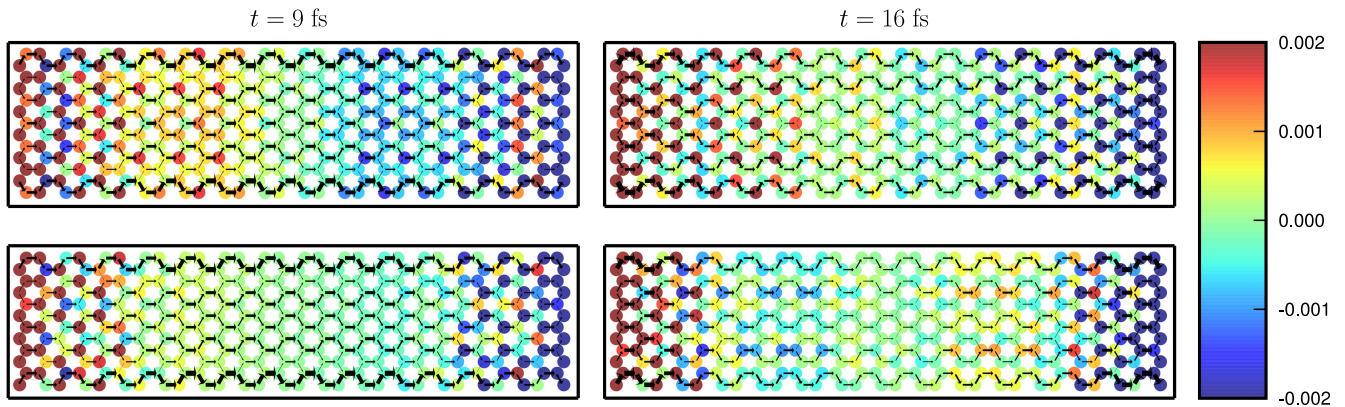}
\caption{Temporal snapshots of spatial charge densities and bond currents along aGNRs.
         Upper panel shows the fully symmetric aGNR and lower panel transersally asymmetric aGNR.
         Left panel shows the snapshots corresponding to the first maximum in the transient current
         and the right panel shows the ones corresponding to the first minimum.
         The charge densities are calculated as the difference from the ground-state
         density (colour map). The bond currents are drawn as solid arrows where the
         width of the arrow indicates the relative strength of the current.}
\label{fig:paritya}
\end{figure*}

\begin{figure*}[ht!]
\includegraphics[width=\textwidth]{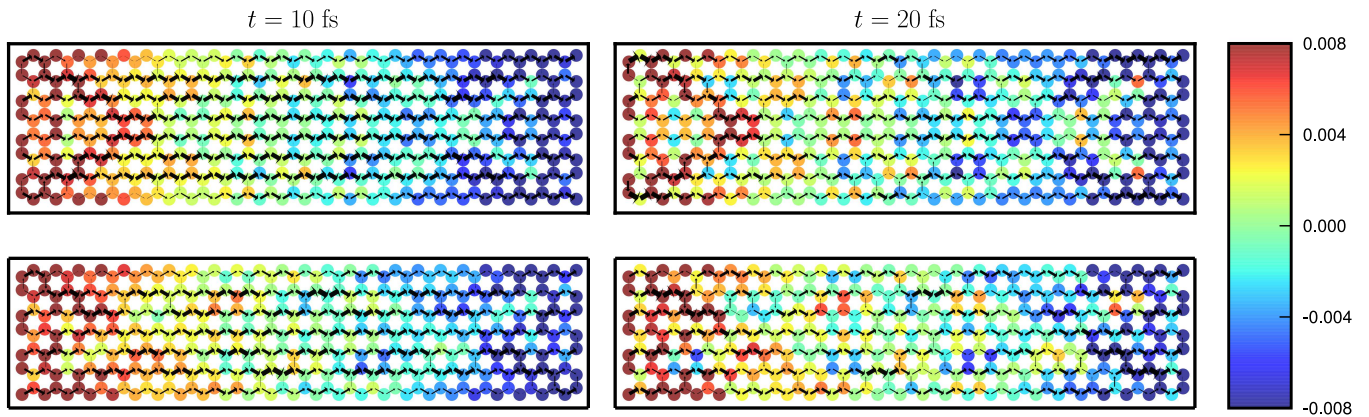}
\caption{Temporal snapshots as in Fig.~\ref{fig:paritya} but for zGNRs and at different times.}
\label{fig:parityz}
\end{figure*}

In Fig.~\ref{fig:length}a we notice the formation of quasi-stationary 
states as we increase the length of the ribbon. The current steeply 
increases from zero to some value and then grows linearly before 
decreasing again. The
growth is slower and lasts longer the longer is the ribbon. 
Let us investigate further the dependence of the current on 
the length of the ribbon. In Fig.~\ref{fig:plateau} we show the 
transient currents through $13$-aGNR ($W=1.4$ nm) and $8$-zGNR ($W=1.6$ nm) of
similar lengths with $V_{\text{sd}}= 5{.}6$ eV. For graphical 
purposes we normalize the current by the number of bonds in the bridge 
and the time by the length $L$ of the ribbon. The curves do 
essentially collapse on one single curve. The peculiar feature of the 
aGNRs is the current plateau for $1\lesssim t/L\lesssim 2$. The 
duration of the plateau corresponds to the time for an electron with 
velocity $v \sim 1$ nm/fs to cross the ribbon. This velocity is consistent 
with the value of the Fermi velocity $v_{\mathrm{F}}=3|t_{C}|a/(2\hbar)$
where $a=1{.}42$ \AA \, is the carbon--carbon distance.\cite{CGPNG.2009} The physical 
picture is that an almost step-like, right-moving density wave reaches the bridge 
(positioned in the middle of the ribbon) at $t/L\simeq 
1/2$ and the right interface at $t/L\simeq 1$. At this time the 
wave is reflected backward and at time $t/L\simeq 3/2$ reaches the bridge 
thus destroying the plateau. No pronounced plateau is instead observed in zGNRs. 
As we shall see in the next Section the current distribution along 
the ribbon is strongly dependent on the orientation of the bonds. The 
tilted bonds in zGNRs cause multiple reflections at the edges thus 
preventing the formation of a current plateau. Also, more powerful reflection can be seen from
the zigzag edge state (at the lead interface) in the case of aGNRs.

\subsection{Even--odd parity effects in charge and current profiles}\label{sec:parity}

\begin{figure}[t]
\includegraphics{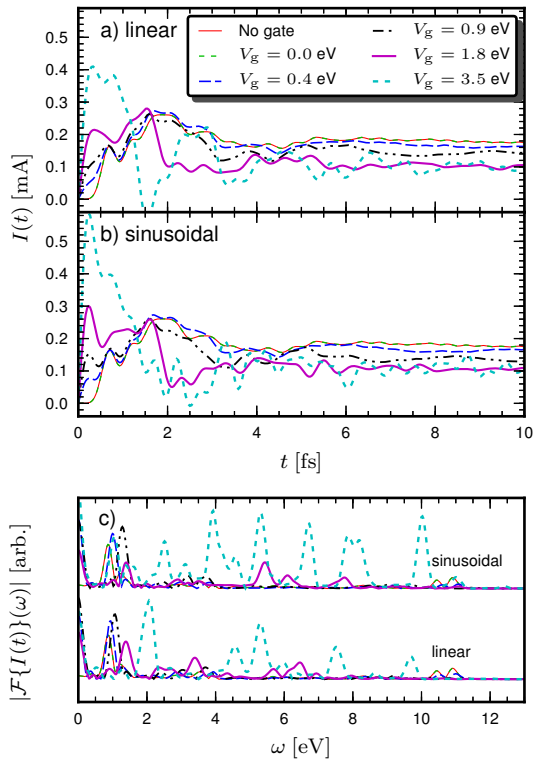}
\caption{Time-dependent bond currents through a $4$-zGNR (length $0{.}7$ nm and width $0{.}9$ nm)
         with fixed bias voltage $V_{\text{sd}}/2=3{.}5$ eV and with 
	 \emph{varying potentials}:
         a) linear potential profile,
         b) sinusoidal potential profile,
         c) the corresponding Fourier transforms (sinusoidal is offset for clarity).}
\label{fig:gate}
\end{figure}

\begin{figure}[t]
\includegraphics[width=0.55\textwidth]{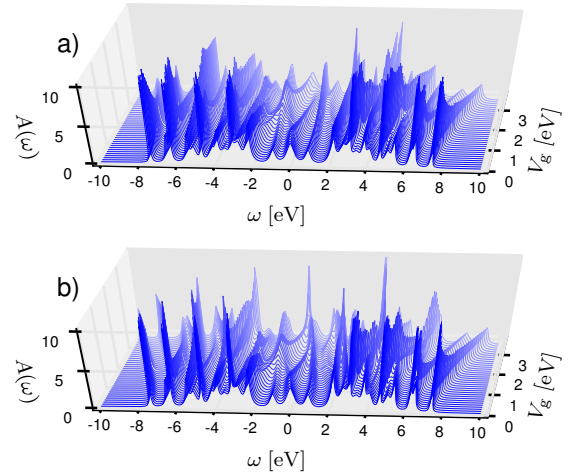}
\caption{Nonequilibrium spectral functions of the studied zGNR with 
\emph{varying potential}:
         a) linear potential profile, and
         b) sinusoidal potential profile.}
\label{fig:spectral}
\end{figure}

The GNRs are parametrized by integer numbers (even or symmetric and 
odd or asymmetric) for width and length. In this Section we study 
how the parity of the GNRs affects the charge 
and current profiles in the transient regime. 
We choose ribbons of equivalent lengths, approximately $6$ nm ($14$ armchair cells and $25$ 
zigzag cells) and equivalent widths, approximately $1{.}5$ nm. However, we take the widths as
$\{7,8\}$ zigzag-lines and $\{12,13\}$ armchair dimer-lines which, in turn, correspond to either 
symmetrical or asymmetrical ribbon in the 
longitudinal direction, see Figs.~\ref{fig:paritya} 
and~\ref{fig:parityz}. A bias voltage $V_{\text{sd}} = 5{.}6$ eV is 
applied to the leads and $V_{\rm g}$ is set to zero.
In Figs.~\ref{fig:paritya} and~\ref{fig:parityz} we show snapshots of 
the density variation and bond-current profiles. 
The density variation is defined as the difference between the 
density at time $t$ and the
ground-state density. Since the size of the ribbons is comparable 
to that in Sec.~\ref{sec:transient}, we choose the snapshot times 
to correspond to the first wavecrest, $\{9,10\}$ fs (on the left panels) and to 
the first wavetrough, $\{16,20\}$ fs (on the right panels). 
The full density and current dynamics is shown in an animation.\cite{animation}

The symmetry of the ribbon is responsible for the charge and current profiles. 
In the aGNR case, see Fig.~\ref{fig:paritya}, the top
panel shows a fully symmetric $13$-aGNR (invariant structure for mirrorings both in the transverse
and longitudinal direction) and the bottom panel shows a $12$-aGNR (invariant structure for
mirroring only in the longitudinal direction). The asymmetry does not
lead to dramatic differences in the charge and current distributions. In the charge profile of the
symmetric aGNR certain `cold' and `hot' spots show up in the middle region whereas in the
asymmetric aGNR the charge is more evenly distributed from the source electrode to the drain
electrode. Also, in both aGNR structures the current is mostly 
flowing through the edges and the wavefront is flat.\cite{animation} In the
zGNR case, see Fig.~\ref{fig:parityz}, the top panel shows an even $8$-zGNR and the bottom panel shows
an odd $7$-zGNR. In both structures we observe diagonal charge patterns along the ribbon; in the even
zGNR these patterns are symmetric whereas in the odd zGNR the patterns show asymmetric features.
Certain `cold' and `hot' spots show up in the crossings of density wavefronts. In addition,
the current is mostly flowing longitudinally through the interior of 
the ribbons with a much smaller contribution coming from the edges.
From the animation in Ref. \onlinecite{animation} we also see that 
the wavefront has a triangular shape.

The pattern of the charge--current profile is quite different at 
different times. On the left panels we have a perfect wave propagating along the
ribbon, whereas on the right panels we see an interference pattern 
due to the reflected wave. 
In the density wave profile there are two antinodes at the electrode interfaces, at the time
corresponding to the first maximum ($t=10$ fs) and one antinode together with two nodes in the 
middle region. At the time corresponding to the first minimum ($t=20$ fs) the antinodes
remain at the electrode interface but additional nodes arise 
in the middle region. 

\subsection{Perturbed central region}

As an illustration of the formula in Eq. (\ref{eq:td-dens-gate}) for perturbed central regions 
we study the transient of a $4$-zGNR (or more accurately a `$4$-by-$4$ graphene flake'). The
system consists of $32$ carbon sites and an on-site potential 
$\phi_{m}$ is 
switched on at site $m$ concurrently with the applied bias.
Let us investigate how the form of the voltage profile within the flake affects the
transient dynamics. We define $x_{m}$ to be the distance of the 
$m$th carbon atom from the left interface and take 
$\phi_{m}=\phi(x_m)$.
For a linear potential profile we use
\be
\phi(x_m) = -\frac{2V_{\text{g}}}{L} \, x_m + V_{\text{g}} \ , \nonumber
\ee
and for a sinusoidal potential profile
\be
\phi(x_m) = \begin{cases} V_{\text{g}} \ , & \ x_m<L/10 \\
                        V_{\text{g}}\cos\left(\frac{5\pi}{4L}x_m-\frac{\pi}{8}\right) \ , &  L/10 \leq x_m \leq 9L/10 \\ 
                        -V_{\text{g}} \ , & x_m>9L/10 \ , \end{cases}\nonumber
\ee
where $L$ is the length of the flake.

In Fig.~\ref{fig:gate} we show the time-dependent currents through the flake with fixed bias
voltage $V_{\text{sd}}/2 = 3{.}5$ eV and varying linear potential in panel a, and 
sinusoidal potential in panel b. The comparison with the previous result
of non-perturbed, `No gate', and perturbed,  `$V_{\text{g}}=0.0$ eV', 
central region provides a numerical check of the correctness of Eq. 
(\ref{eq:td-dens-gate}). 

For voltages smaller than $1$ eV the transient is not so 
different from the non-perturbed results. However, for stronger voltages a rather
non-trivial transient behaviour is observed. Notice that the largest 
value $V_{\text{g}}=3.5$ eV corresponds to the physical situation of 
a continuous potential profile. 
The Fourier spectrum of the transient is shown in panel c. The much 
richer structure in several high-energy spectral windows is due to 
transitions involving levels of the {\em perturbed} central region.

The dependence of the energy and spectral weigth of the levels on 
$V_{\rm g}$ is most clearly visualized by plotting 
the nonequilibrium spectral function
\be\label{eq:spectral}
A(\w) = -\frac{1}{\pi}\Im\mathrm{Tr}\left[G^{\mathrm{R}}(\w)\right] 
\ee
where the trace is over the states of the central region. 
The spectral function is displayed in Fig.~\ref{fig:spectral}.
As expected the spectrum widens with increasing $V_{\text{g}}$.
The high energy peaks at $\w \approx \pm8$ eV 
(in the non-perturbed case: $V_{\text{g}}=0$ eV) shift to $\w \approx \pm 10$ eV 
(when the perturbation is at its maximum: $V_{\text{g}}=3{.}5$ eV). This 
is consistent with the peaks occuring at around $\w \approx 10$ eV
in Fig. \ref{fig:gate}c. With a similar analysis one can show that all other main 
peaks in the Fourier spectrum can be interpreted by inspecting the spectral function.


\section{Conclusion}\label{sec:conclusion}
In this work we developed a time-dependent extension of the Landauer--B\"uttiker approach to study 
transient dynamics in time-dependent quantum transport through molecular junctions. We have derived 
a closed integral expression for the time-dependence of the density matrix of the molecular junction
after switch-on of a bias voltage in the leads or a perturbation in the junction as well as for the
current flowing into the leads. Both equations can be evaluated without the necessity of
propagating individual single-particle orbitals or Green's functions.
We applied the approach to study the transient dynamics of zigzag and armchair graphene nanoribbons 
of different symmetries. We found a rich transient dynamics in which the saturation times can exceed
several hundreds of femtoseconds while displaying a long time oscillatory motion related to multiple
reflections of the density wave in the nanoribbons at the ribbon--lead interface. In the case of 
armchair nanoribbons we find pronounced quasi-steady states which can be explained by multiple 
reflections of the density wave passing through the ribbon with the edge states located at the 
ribbon--lead interfaces. We see further in the case of zigzag nanoribbons that there is a 
predominant oscillation frequency associated with virtual transitions between the edge states and 
the Fermi levels of the electrode. The transient dynamics therefore give detailed spectral 
information on the structure of the nanoribbons. Recently the ultrafast dynamics of individual 
carbon nanotubes has been measured using laser optics by four-wave mixing techniques.\cite{petterson} 
There are therefore important experimental 
developments that can, in the future, give access to the direct study of transient dynamics. Such 
transient spectroscopy can give important detailed informations on the structure of molecular 
junctions out of equilibrium.


\acknowledgments
RT wishes to thank V{\"a}is{\"a}l{\"a} Foundation of The Finnish Academy of Science and Letters 
for financial support and CSC --- the Finnish IT Center for Science --- for computing resources. 
RvL thanks the Academy of Finland for support. EP and GS acknowledge funding by MIUR FIRB Grant
No. RBFR12SW0J. GS acknowledges financial support through travel grant Psi-K2 5813 
   of the European Science Foundation (ESF).
C. Gomes da Rocha, A.-M. Uimonen, N. S{\"a}kkinen 
and M. Hyrk{\"a}s are acknowledged for useful discussions.


\appendix
\section{Results in the zero-temperature limit}\label{app:formulae}

By taking into account the behaviour of the Fermi function in the zero-temperature
limit and adjusting accordingly the integrals in Eqs.~\eqref{eq:terms1}, 
\eqref{eq:terms2} and~\eqref{eq:terms3}, we get the following explicit expressions 
\begin{widetext}
\beq
\itL_{\a,jk} & = & \frac
                   {\mathrm{Log}(\eps_k^*-\mu_\a) - \mathrm{Log}(\eps_j-\mu_\a)}
                   {2\pi(\eps_k^* - \eps_j)} \ ,
                   \label{eq:lambda} \\
\itP_{\a,jk}(t) & = & \frac{\ex^{-\im (\eps_j-\mu_\a) t}\left\{ F[\im (\eps_k^* - \mu_\a) t]
                            + \frac{\eps_k^* - \eps_j - V_{\a}}{V_{\a}}F[\im (\eps_j - \mu_\a) t]
                            - \frac{\eps_k^* - \eps_j}{V_{\a}}F[\im (\eps_j - \mu) t]\right\}}
                           {2\pi (\eps_k^* - \eps_j) (\eps_k^* - \eps_j - V_{\a})} \ ,
                            \label{eq:pi} \\
\itO_{\a,jk} & = & \frac{(\eps_k^* - \eps_j + V_{\a})\left[\mathrm{Log}(\eps_k^* - \mu_\a) - \mathrm{Log}(\eps_j - \mu)\right] + (\eps_k^* - \eps_j - V_{\a})\left[\mathrm{Log}(\eps_j - \mu_\a) - \mathrm{Log}(\eps_k^* - \mu)\right]}{2\pi\left[(\eps_k^* - \eps_j)V_{\a}^3 - (\eps_k^* - \eps_j)^3 V_{\a}\right]}
\label{eq:omega}
\eeq
\end{widetext}
where we defined $\mu_\a = \mu + V_{\a}$ and
\be\label{eq:f}
F(z) = \begin{cases}\ex^z\left[2\pi\im - \mathrm{E}_{1}(z)\right] \ , & \text{if} \ \mathrm{Arg}(z) \in ]-\pi,-\pi/2] \\ -\ex^z\mathrm{E}_{1}(z) \ , & \text{otherwise.}\end{cases} 
\ee
$\mathrm{Log}$ is the principal branch complex logarithm function, $\mathrm{Arg}$ 
the principal argument and $\mathrm{E}_{1}$ 
the exponential integral function:
\be\label{eq:expint}
\mathrm{E}_{1}(z) = \int_1^\infty \frac{\ex^{-z t}}{t} \ \ud t \ .
\ee
About the implementation of the complex-valued (complex variable) exponential integral 
there is a thorough introduction in Ref.~\onlinecite{expint}.
The piecewise definition of the function $F$ is due to branch cuts in the $z$-plane.

We notice in Eqs.~\eqref{eq:lambda}, \eqref{eq:pi} and~\eqref{eq:omega} that it is possible 
that the structure of the single-particle Hamiltonian $h$ would together with the 
coupling matrices $\itG$ produce such an effective Hamiltonian $h_{\mathrm{eff}}$ with 
degenerate eigenvalues: $\Im \eps_j = 0$ and $\Re \eps_j = \Re \eps_k^*$.
In this case we consider the 
left/right eigenbasis of the effective Hamiltonian $h_{\mathrm{eff}}$:  Since 
$h_{\mathrm{eff}} = h - \frac{\im}{2}\itG$, where $h$ and $\itG$ are hermitian matrices, then
\beq
\eps_j\ev{\Psi_j^{\text{L}}}{\Psi_j^{\text{L}}} & = & \bra{\Psi_j^{\text{L}}}h_{\mathrm{eff}}\ket{\Psi_j^{\text{L}}} \nonumber \\
& = & \bra{\Psi_j^{\text{L}}}h\ket{\Psi_j^{\text{L}}} - \frac{\im}{2}\bra{\Psi_j^{\text{L}}}\itG\ket{\Psi_j^{\text{L}}} 
\eeq
which, in turn, gives
\be
\eps_j = \frac{\bra{\Psi_j^{\mathrm{L}}}h\ket{\Psi_j^{\mathrm{L}}}}{\ev{\Psi_j^{\text{L}}}{\Psi_j^{\text{L}}}}
       - \frac{\frac{\im}{2}\bra{\Psi_j^{\mathrm{L}}}\itG\ket{\Psi_j^{\mathrm{L}}}}{\ev{\Psi_j^{\text{L}}}{\Psi_j^{\text{L}}}} \ .
\ee
Since the expectation values are real and $\itG$ is a positive definite matrix, we get
\be
\Im \eps_j = - \frac{1}{2}\frac{\bra{\Psi_j^{\mathrm{L}}}\itG\ket{\Psi_j^{\mathrm{L}}}}{\ev{\Psi_j^{\text{L}}}{\Psi_j^{\text{L}}}} < 0 \ .
\ee
Then suppose that $\Im \eps_j = 0$. This gives $\bra{\Psi_j^{\mathrm{L}}}\itG\ket{\Psi_j^{\mathrm{L}}} = 0$, and since the 
level-width matrices are calculated from the tunneling matrices by $\itG \sim T^\dagger T$, we get
\be
\bra{\Psi_j^{\mathrm{L}}}T^\dagger T\ket{\Psi_j^{\mathrm{L}}} = 0 \ \Rightarrow \ \ev{\chi_j^{\text{L}}}{\chi_j^{\text{L}}} = 0 \ ,
\ee
where $\ket{\chi_j^{\mathrm{L}}} = T \ket{\Psi_j^{\mathrm{L}}}$. Having then a zero-norm vector $\ket{\chi_j^{\mathrm{L}}}$ it means that 
vector itself must be zero, i.e., $0 = \ket{\chi_j^{\mathrm{L}}} = T\ket{\Psi_j^{\mathrm{L}}}$ for all $j$. This means 
that $\ket{\Psi_j^{\mathrm{L}}}$ is an eigenvector of $T$ with zero eigenvalue. In particular 
$\itG \ket{\Psi_j^{\mathrm{L}}} = T^\dagger T\ket{\Psi_j^{\mathrm{L}}} = 0$, and hence
\be
\itG_{jk} = \mel{\Psi_{j}^{\text{L}}}{\itG}{\Psi_{k}^{\text{L}}} = 0 \ , \ \forall j,k \ .
\ee
Therefore the case of degenerate eigenvalues can be excluded from the derived formulae all together.
This also relates to some particular systems having states that are eigenfunctions of 
$\itG_{\a,mn}$ with zero eigenvalue. In these cases, it 
becomes important to take into account the infinitesimal $\im \eta$ in the retarded 
Green's function for these states, i.e. the Green's function operator acting on 
these states has the effective form $G^\textrm{R} (\omega) = (\omega - h + \im\eta)^{-1}$.
 This effectively amounts to an infinitesimal value of $\itG_{\a,mn}$ for these particular states
in Eq.~\eqref{eq:td-dens-decomposed} which leads to sharp delta peaks in the spectral function.  
However, since these states are inert and do not contribute to the dynamics they only affect the static 
part of the density matrix. Numerically it is then more advantageous to calculate these states separately 
and add a cut-off in Eq.~\eqref{eq:td-dens-decomposed}. We evaluate Eq.~\eqref{eq:td-dens-decomposed} 
only for $\itG_{\a,mn} > \epsilon$ with $\epsilon$ a small number and treat the inert states separately. 
The part of the density matrix corresponding to these inert states is then given by
\begin{equation}
\hat{\rho} = \sum_{\epsilon_j < \mu} |\phi_j \rangle \langle \phi_j |   
\end{equation}
where we sum over all eigenstates of $h$ that satisfy $\itG_{\alpha} | \phi_j \rangle=0$ for all $\alpha$.
Note that the existence of the inert states is a very special case caused by symmetries of the
molecule and $\itG_\a$.  The only case we encountered in the present study where such states exist 
is the case of the fully symmetric aGNR of Fig.~\ref{fig:paritya}. There the inert states are given 
by wave functions that have nodal planes exactly at the rows which are contacted to the leads. 

\section{Results for the perturbed central region}\label{app:gate}
In this Appendix we guide the reader through the derivation of 
Eq.~\eqref{eq:td-dens-gate}. As we will often 
refer to results in Ref.~\onlinecite{rikuproc} we here append the 
suffix ``I'' to every equation or section in this reference.

The results of Ref. ~\onlinecite{rikuproc} are general and remain valid  in the 
presence of electric or magnetic fields in the central region until Sec.~3.2-I.
In the Green's function calculations the Matsubara Green's function 
does not change as it depends only on 
the ground-state Hamiltonian $h_{CC}$. On the other hand, for Green's functions
having components on the horizontal branch of the Keldysh contour
we have to use the Hamiltonian $\widetilde{h}_{CC}$. 
Therefore the Eqs.~(24-I) and~(25-I) change according to
\begin{widetext}
\begin{eqnarray}
G^\rceil(t,\tau) & = & \ex^{-\im \widetilde{h}_{\text{eff}} t}\left[G^{\text{M}}(0,\tau)-\int_0^t\mathrm{d} t' \mathrm{e}^{\mathrm{i}\widetilde{h}_{\text{eff}} t'}\int_0^\beta\mathrm{d} \bar{\tau} \itS^\rceil(t',\bar{\tau})G^{\text{M}}(\bar{\tau},\tau) \right] \ , \\
G^{\text{R}}(t-t') & = & -\im \theta(t-t')\mathrm{e}^{-\im\widetilde{h}_{\text{eff}} (t-t')} 
\end{eqnarray}
\end{widetext}
where $\widetilde{h}_{\text{eff}} = \widetilde{h}_{CC}-\frac{\mathrm{i}}{2}\itG$.
All steps in Appendix C-I and D-I as well as in Sec.~3.3-I
should change accordingly. In particular we stress the  $G^{\text{M}}$ and 
$G^{\text{R}}$ in Eq.~(C.9-I) are now different, that 
$\widetilde{V}_\alpha$ is a matrix (Appendix D-I) and that the 
Dyson-like equation [Eq.~(D.1-I)] relating the non-perturbed and perturbed Green's 
functions now reads
\begin{equation}
G^{\text{R}}(\omega)-\widetilde{G}^{\text{R}}(\omega+V_\alpha) = G^{\text{R}}(\omega)\widetilde{V}_\alpha\widetilde{G}^{\text{R}}(\omega+V_\alpha) \ .
\end{equation}
With these considerations and following the same steps as in Ref.~\onlinecite{rikuproc}
we arrive at the result shown in Eq.~\eqref{eq:td-dens-gate}.

Next, by expanding in the \emph{left} eigenbasis of $\widetilde{h}_{\mathrm{eff}}$ 
we find 
\begin{widetext}
\be\label{eq:td-dens-decomposed-gate}
\widetilde{\rho}_{jk}(t) = \mel{\widetilde{\Psi}_j^{\text{L}}}{\rho(t)}{\widetilde{\Psi}_k^{\text{L}}} = \sum_\a\left[\widetilde{\itG}_{\a,jk}\widetilde{\itL}_{\a,jk} + \widetilde{\itP}_{\a,jk}(t) + \widetilde{\itP}_{\a,kj}^*(t) + \widetilde{\itO}_{\a,jk}(t)\right] 
\ee
with the introduced functions
\beq\label{eq:terms-gate}
\widetilde{\itG}_{\a,jk} & = & \mel{\widetilde{\Psi}_j^{\text{L}}}{\itG_{\a}}{\widetilde{\Psi}_k^{\text{L}}} \ , \nonumber\\
\widetilde{\itL}_{\a,jk} & = & \intw\frac{f(\omega - \mu)}{(\w+V_{\a}-\widetilde{\eps}_j)(\w+V_{\a}-\widetilde{\eps}_k^*)} \ , \nonumber\\ 
\widetilde{\itP}_{\a,jk}(t) & = & \sum\limits_{m,n}\frac{\ev{\widetilde{\Psi}_j^{\text{L}}}{\Psi_m^{\text{R}}}\mel{\Psi_m^{\text{L}}}{\widetilde{V}_{\a}}{\widetilde{\Psi}_n^{\text{R}}}\widetilde{\itG}_{\a,nk}}{\ev{\Psi_m^{\text{L}}}{\Psi_m^{\text{R}}}\ev{\widetilde{\Psi}_n^{\text{L}}}{\widetilde{\Psi}_n^{\text{R}}}} \intw\frac{f(\omega - \mu)\ex^{\im(\w+V_{\a}-\widetilde{\eps}_j)t}}{(\w-\eps_m)(\w+V_{\a}-\widetilde{\eps}_n)(\w+V_{\a}-\widetilde{\eps}_k^*)} \ , \nonumber\\ 
\widetilde{\itO}_{\a,jk}(t) & = & \sum\limits_{m,n,p,q}\frac{\ev{\widetilde{\Psi}_j^{\text{L}}}{\Psi_m^{\text{R}}}\mel{\Psi_m^{\text{L}}}{\widetilde{V}_{\a}}{\widetilde{\Psi}_n^{\text{R}}}\widetilde{\itG}_{\a,np}\mel{\widetilde{\Psi}_p^{\text{R}}}{\widetilde{V}_{\a}^\dagger}{\Psi_q^{\text{L}}}\ev{\Psi_q^{\text{R}}}{\widetilde{\Psi}_k^{\text{L}}}}{\ev{\Psi_m^{\text{L}}}{\Psi_m^{\text{R}}}\ev{\widetilde{\Psi}_n^{\text{L}}}{\widetilde{\Psi}_n^{\text{R}}}\ev{\widetilde{\Psi}_p^{\text{R}}}{\widetilde{\Psi}_p^{\text{L}}}\ev{\Psi_q^{\text{R}}}{\Psi_q^{\text{L}}}} \nonumber \\
& \times & \ex^{-\im(\widetilde{\eps}_j-\widetilde{\eps}_k^*)t}\intw\frac{f(\omega - \mu)}{(\w-\eps_m)(\w+V_{\a}-\widetilde{\eps}_n)(\w+V_{\a}-\widetilde{\eps}_p^*)(\w-\eps_q^*)}  \nonumber\\
\eeq
where eigenvalues $\eps_j$ and $\widetilde{\eps}_k^*$ refer to the complex eigenvalues of 
$h_{\mathrm{eff}}$ and $\widetilde{h}_{\mathrm{eff}}$, respectively. 
In the limit $\widetilde{h}_{\mathrm{eff}} \to h_{\mathrm{eff}}$ this result can 
also be checked to reduce to the earlier result in Eqs.~\eqref{eq:td-dens-decomposed}, \eqref{eq:terms1}, \eqref{eq:terms2} and~\eqref{eq:terms3}. 
In the limit of uncontacted system Eq.~\eqref{eq:td-dens-decomposed-gate} describes the dynamics of an isolated (perturbed) system,
in which case the same result could be derived directly from the equations of motion of the
one-particle density matrix.

In the zero-temperature limit, the integrals 
in Eq.~\eqref{eq:terms-gate} can be calculated analytically also in this case. The integrals 
now only have more constants and the final results can not be simplified as much as earlier. 
The explicit forms can be found below
\beq
\widetilde{\itL}_{\a,jk} & = & \frac{\mathrm{Log}(\widetilde{\eps}_k^*-\mu_\a)-\mathrm{Log}(\widetilde{\eps}_j-\mu_\a)}{2\pi(\widetilde{\eps}_k^*-\widetilde{\eps}_j)} \ , \label{eq:gate-lambda}\\ 
\widetilde{\itP}_{\a,jk}(t) & = & \sum_{m,n}\frac{\ev{\widetilde{\Psi}_j^{\text{L}}}{\Psi_m^{\text{R}}}\mel{\Psi_m^{\text{L}}}{\widetilde{V}_{\a}}{\widetilde{\Psi}_n^{\text{R}}}\mel{\widetilde{\Psi}_n^{\text{L}}}{\itG_{\a}}{\widetilde{\Psi}_k^{\text{L}}}}{\ev{\Psi_m^{\text{L}}}{\Psi_m^{\text{R}}}\ev{\widetilde{\Psi}_n^{\text{L}}}{\widetilde{\Psi}_n^{\text{R}}}}\frac{\ex^{-\im(\widetilde{\eps}_j-\mu_\a)t}}{2\pi(\widetilde{\eps}_k^*-\widetilde{\eps}_n)(\widetilde{\eps}_k^*-\eps_m-V_{\a})} \nonumber \\
&  & \times \ \left\{F[\im(\widetilde{\eps}_k^*-\mu_\a)t] - \frac{\widetilde{\eps}_k^*-\eps_m-V_{\a}}{\widetilde{\eps}_n-\eps_m-V_{\a}}F[\im(\widetilde{\eps}_n-\mu_\a)t]+\frac{\widetilde{\eps}_k^*-\widetilde{\eps}_n}{\widetilde{\eps}_n-\eps_m-V_{\a}}F[\im(\eps_m-\mu)t]\right\} \ , \label{eq:gate-pi}\\ 
\widetilde{\itO}_{\a,jk}(t) & = & \sum_{m,n,p,q}\frac{\ev{\widetilde{\Psi}_j^{\text{L}}}{\Psi_m^{\text{R}}}\mel{\Psi_m^{\text{L}}}{\widetilde{V}_{\a}}{\widetilde{\Psi}_n^{\text{R}}}\mel{\widetilde{\Psi}_n^{\text{L}}}{\itG_{\a}}{\widetilde{\Psi}_p^{\text{L}}}\mel{\widetilde{\Psi}_p^{\text{R}}}{\widetilde{V}_{\a}^\dagger}{\Psi_q^{\text{L}}}\ev{\Psi_q^{\text{R}}}{\widetilde{\Psi}_k^{\text{L}}}}{\ev{\Psi_m^{\text{L}}}{\Psi_m^{\text{R}}}\ev{\widetilde{\Psi}_n^{\text{L}}}{\widetilde{\Psi}_n^{\text{R}}}\ev{\widetilde{\Psi}_p^{\text{R}}}{\widetilde{\Psi}_p^{\text{L}}}\ev{\Psi_q^{\text{R}}}{\Psi_q^{\text{L}}}} \frac{\ex^{-\im(\widetilde{\eps}_j-\widetilde{\eps}_k^*)t}}{2\pi} \nonumber \\
& & \times \left[ \frac{\mathrm{Log}(\eps_m-\mu)}{(\eps_m-\widetilde{\eps}_n+V_{\a})(\eps_m-\widetilde{\eps}_p^*+V_{\a})(\eps_m-\eps_q^*)} + \frac{\mathrm{Log}(\widetilde{\eps}_n-\mu_\a)}{(\widetilde{\eps}_n-\eps_m-V_{\a})(\widetilde{\eps}_n-\widetilde{\eps}_p^*)(\widetilde{\eps}_n-\eps_q^*-V_{\a})} \right. \nonumber \\
& & \left. +\frac{\mathrm{Log}(\eps_q^*-\mu)}{(\eps_q^*-\eps_m)(\eps_q^*-\widetilde{\eps}_n+V_{\a})(\eps_q^*-\widetilde{\eps}_p^*+V_{\a})} + \frac{\mathrm{Log}(\widetilde{\eps}_p^*-\mu_\a)}{(\widetilde{\eps}_p^*-\eps_m-V_{\a})(\widetilde{\eps}_p^*-\widetilde{\eps}_n)(\widetilde{\eps}_p^*-\eps_q^*-V_{\a})} \right] \label{eq:gate-omega}
\eeq
where $\mu_\a = \mu + V_{\a}$ and $F$ is as in Eq.~\eqref{eq:f}. Also these results can be 
checked to reduce to the earlier results in 
Eqs.~\eqref{eq:lambda}, \eqref{eq:pi} and~\eqref{eq:omega} when $\widetilde{\Psi}\to\Psi$ and
$\widetilde{\eps} \to \eps$ ($\widetilde{h}_{\mathrm{eff}} \to h_{\mathrm{eff}}$).
\end{widetext}





\begin{thebibliography}{99}

\bibitem{landauer}
R.~Landauer,
\href{http://ieeexplore.ieee.org/Xplore/defdeny.jsp?url=http%3A%2F%2Fieeexplore.ieee.org%2Fstamp%2Fstamp.jsp%3Ftp%3D%26arnumber%3D5392683&denyReason=-133&arnumber=5392683&productsMatched=null&userType=inst}{IBM J. Res. Dev. {\bf 1}, 223 (1957).}

\bibitem{buttiker}
M.~B{\"u}ttiker,
\href{http://link.aps.org/doi/10.1103/PhysRevLett.57.1761}{ Phys. Rev.  Lett. {\bf 57}, 1761 (1986).}

\bibitem{L.1995}
N.~D. Lang,
\href{http://link.aps.org/doi/10.1103/PhysRevB.52.5335}{Phys. Rev. B {\bf 52}, 5335 (1995).}

\bibitem{LA.1998}
N.~D. Lang and P.~Avouris,
\href{http://link.aps.org/doi/10.1103/PhysRevLett.81.3515}{Phys. Rev. Lett. {\bf 81}, 3515 (1998).}

\bibitem{TGW-1.2001}
J.~Taylor, H.~Guo, and J.~Wang,
\href{http://link.aps.org/doi/10.1103/PhysRevB.63.121104}{Phys. Rev. B {\bf 63}, 121104  (2001).}

\bibitem{TGW-2.2001}
J.~Taylor, H.~Guo, and J.~Wang.
\href{http://link.aps.org/doi/10.1103/PhysRevB.63.245407}{Phys. Rev. B {\bf 63}, 245407  (2001).}

\bibitem{BMOTS.2002}
M.~Brandbyge, J.~L. Mozos, P.~Ordejon, J.~Taylor, and K.~Stokbro,
\href{http://link.aps.org/doi/10.1103/PhysRevB.65.165401}{Phys. Rev. B {\bf 65}, 165401 (2002).}

\bibitem{SA.2004}
G.~Stefanucci and C.-O. Almbladh,
\href{http://iopscience.iop.org/0295-5075/67/1/014/}{Europhys. Lett. {\bf 67}, 14 (2004).}

\bibitem{FWK.2004}
F.~Evers, F.~Weigend, and M.~Koentopp,
\href{http://prb.aps.org/abstract/PRB/v69/i23/e235411}{Phys. Rev. B {\bf 69}, 235411 (2004).}

\bibitem{KSARG.2005}
S.~Kurth, G.~Stefanucci, C.-O. Almbladh, A.~Rubio, and E.~K.~U. Gross,
\href{http://link.aps.org/doi/10.1103/PhysRevB.72.035308}{Phys. Rev. B {\bf 72}, 035308 (2005).}

\bibitem{QLLY.2006}
X.~Qian, J.~Li, X.~Lin, and S.~Yip,
\href{http://link.aps.org/doi/10.1103/PhysRevB.73.035408}{Phys. Rev. B {\bf 73}, 035408 (2006).}

\bibitem{SPC.2008}
G.~Stefanucci, E.~Perfetto, and M.~Cini,
\href{http://link.aps.org/doi/10.1103/PhysRevB.78.075425}{Phys. Rev. B {\bf 78}, 075425 (2008).}

\bibitem{ZC.2009}
Z.~Zhou and S.-I. Chu,
\href{http://iopscience.iop.org/0295-5075/88/1/17008/}{Europhys. Lett. {\bf 88}, 17008 (2009).}

\bibitem{SPC.2010}
G.~Stefanucci, E.~Perfetto, and M.~Cini,
\href{http://link.aps.org/doi/10.1103/PhysRevB.81.115446}{Phys. Rev. B {\bf 81}, 115446 (2010).}

\bibitem{GWSHGW.2014}
B.~Gaury et~al.,
\href{http://www.sciencedirect.com/science/article/pii/S0370157313003451}{Phys. Rep. {\bf 534}, 1 (2014) }

\bibitem{BCG.2008}
P.~Bokes, F.~Corsetti, and R.~W. Godby,
\href{http://link.aps.org/doi/10.1103/PhysRevLett.101.046402}{Phys. Rev. Lett. {\bf 101}, 046402 (2008).}

\bibitem{CFPS.2009}
A.~Chaves, G.~A. Farias, F.~M. Peeters, and B.~Szafran,
\href{http://link.aps.org/doi/10.1103/PhysRevB.80.125331}{Phys. Rev. B {\bf 80}, 125331 (2009).}

\bibitem{B.2011}
P.~Bokes,
\href{http://link.aps.org/doi/10.1103/PhysRevA.83.032104}{Phys. Rev. A {\bf 83}, 032104 (2011).}

\bibitem{BSdV.2005}
N.~Bushong, N.~Sai, and M.~D. Ventra,
\href{http://pubs.acs.org/doi/abs/10.1021/nl0520157}{Nano Lett {\bf 5}, 2569 (2005).}

\bibitem{CEvV.2006}
C.-L. Cheng, J.~S. Evans, and T.~V. Voorhis,
\href{http://link.aps.org/doi/10.1103/PhysRevB.74.155112}{Phys. Rev. B {\bf 74}, 155112 (2006).}

\bibitem{SBHdV.2007}
N.~Sai, N.~Bushong, R.~Hatcher, and M.~DiVentra,
\href{http://link.aps.org/doi/10.1103/PhysRevB.75.115410}{Phys. Rev. B {\bf 75}, 115410 (2007).}

\bibitem{EvV.2009}
H.~Eshuis and T.~van Voorhis,
\href{http://pubs.rsc.org/en/content/articlelanding/2009/cp/b912085h#!divAbstract}{Chem. Phys. Phys. Chem. {\bf 11}, 10293 (2009).}

\bibitem{BSIN.2004}
R.~Baer, T.~Seideman, S.~Ilani, and D.~Neuhauser,
\href{http://dx.doi.org/10.1063/1.1640611}{J. Chem. Phys. {\bf 120}, 3387 (2004).}

\bibitem{V.2011}
K.~Varga,
\href{http://link.aps.org/doi/10.1103/PhysRevB.83.195130}{Phys. Rev. B {\bf 83}, 195130 (2011).}

\bibitem{ZCW.2013}
L.~Zhang, J.~Chen, and J.~Wang,
\href{http://link.aps.org/doi/10.1103/PhysRevB.87.205401}{Phys. Rev. B {\bf 87}, 205401 (2013).}

\bibitem{MWG.2006}
J.~Maciejko, J.~Wang, and H.~Guo,
\href{http://link.aps.org/doi/10.1103/PhysRevB.74.085324}{Phys. Rev. B {\bf 74}, 085324 (2006).}

\bibitem{LY.2007}
X.-Q. Li and Y.~J. Yan,
\href{http://link.aps.org/doi/10.1103/PhysRevB.75.075114}{Phys. Rev. B {\bf 75}, 075114 (2007).}

\bibitem{MGM.2007}
V.~Moldoveanu, V.~Gudmundsson, and A.~Manolescu,
\href{http://link.aps.org/doi/10.1103/PhysRevB.76.085330}{Phys. Rev. B {\bf 76}, 085330 (2007).}

\bibitem{PD.2008}
A.~Prociuk and B.~D. Dunietz,
\href{http://link.aps.org/doi/10.1103/PhysRevB.78.165112}{Phys. Rev. B {\bf 78}, 165112 (2008).}

\bibitem{CS.2009}
A.~Croy and U.~Saalmann,
\href{http://link.aps.org/doi/10.1103/PhysRevB.80.245311}{Phys. Rev. B {\bf 80}, 245311 (2009).}

\bibitem{ZCMKTYY.2010}
X.~Zheng et~al.,
\href{http://dx.doi.org/10.1063/1.3475566}{J. Chem. Phys. {\bf 133}, 114101 (2010).}

\bibitem{XWW.2010}
Y.~Xing, B.~Wang, and J.~Wang,
\href{http://link.aps.org/doi/10.1103/PhysRevB.82.205112}{Phys. Rev. B {\bf 82}, 205112 (2010).}

\bibitem{XJTetal.2012}
H.~Xie et~al.,
\href{http://dx.doi.org/10.1063/1.4737864}{J. Chem Phys. {\bf 137}, 044113 (2012).}

\bibitem{ZXW.2012}
L.~Zhang, Y.~Xing, and J.~Wang,
\href{http://link.aps.org/doi/10.1103/PhysRevB.86.155438}{Phys. Rev. B {\bf 86}, 155438 (2012).}

\bibitem{PSS.2013}
A.~Pertsova, M.~Stamenova, and S.~Sanvito,
\href{http://dx.doi.org/10.1088/0953-8984/25/10/105501}{J. Phys.: Cond. Matt. {\bf 25}, 105501 (2013).}

\bibitem{ZCC.2013}
Y.~Zhang, S.~Chen, and G.~H.~Chen,
\href{http://link.aps.org/doi/10.1103/PhysRevB.87.085110}{Phys. Rev. B {\bf 87}, 085110 (2013).}

\bibitem{mceniry}
E.~McEniry et~al.,
\href{http://dx.doi.org/10.1088/0953-8984/19/19/196201}{J. Phys.: Condens. Matter {\bf 19}, 196201 (2007).}

\bibitem{mssvl.2008}
P.~My{\"o}h{\"a}nen, A.~Stan, G.~Stefanucci, and R.~van Leeuwen,
\href{http://iopscience.iop.org/0295-5075/84/6/67001/}{Europhys. Lett. {\bf 84}, 67001 (2008).}

\bibitem{petriprb}
P.~My{\"o}h{\"a}nen, A.~Stan, G.~Stefanucci, and R.~van Leeuwen,
\href{http://link.aps.org/doi/10.1103/PhysRevB.80.115107}{Phys. Rev. B {\bf 80}, 115107 (2009).}

\bibitem{friesen2}
M.~Puig von Friesen, C.~Verdozzi, and C.-O. Almbladh,
\href{http://link.aps.org/doi/10.1103/PhysRevB.82.155108}{Phys. Rev. B {\bf 82}, 155108 (2010).}

\bibitem{PW.2005}
J.~N. Pedersen and A.~Wacker,
\href{http://link.aps.org/doi/10.1103/PhysRevB.72.195330}{Phys. Rev. B {\bf 72}, 195330 (2005).}

\bibitem{WSK.2006}
S.~Welack, M.~Schreiber, and U.~Kleinekath{\"o}fer,
\href{http://dx.doi.org/10.1063/1.2162537}{J. Chem. Phys. {\bf 124}, 044712 (2006).}

\bibitem{WT.2009}
H.~Wang and M.~Thoss,
\href{http://dx.doi.org/10.1063/1.3173823}{J. Chem. Phys. {\bf 131}, 024114 (2009).}

\bibitem{MGM.2009}
V.~Moldoveanu, V.~Gudmundsson, and A.~Manolescu,
\href{http://iopscience.iop.org/1367-2630/11/7/073019/}{New J. Phys. {\bf 11}, 073019 (2009).}

\bibitem{ZSKEB.2009}
P.~Zedler, G.~Schaller, G.~Kiesslich, C.~Emary, and T.~Brandes,
\href{http://link.aps.org/doi/10.1103/PhysRevB.80.045309}{Phys. Rev. B {\bf 80}, 045309 (2009).}

\bibitem{PW.2010}
J.~N. Pedersen and A.~Wacker,
\href{http://www.sciencedirect.com/science/article/pii/S1386947709002550}{Physica E {\bf 42}, 595 (2010).}

\bibitem{WZJY.2013}
S.~Wang, X.~Zheng, J.~Jin, and Y.~J.~Yan,
\href{http://link.aps.org/doi/10.1103/PhysRevB.88.035129}{Phys. Rev. B {\bf 88}, 035129 (2013).}

\bibitem{WETE.2008}
S.~Weiss, J.~Eckel, M.~Thorwart, and R.~Egger,
\href{http://link.aps.org/doi/10.1103/PhysRevB.77.195316}{Phys. Rev. B {\bf 77}, 195316 (2008).}

\bibitem{SF.2009}
M.~Schiro and M.~Fabrizio,
\href{http://link.aps.org/doi/10.1103/PhysRevB.79.153302}{Phys. Rev. B {\bf 79}, 153302 (2009).}

\bibitem{SMR.2010}
D.~Segal, A.~J. Millis, and D.~R. Reichman,
\href{http://link.aps.org/doi/10.1103/PhysRevB.82.205323}{Phys. Rev. B {\bf 82}, 205323 (2010).}

\bibitem{CR.2011}
G.~Cohen and E.~Rabani,
\href{http://link.aps.org/doi/10.1103/PhysRevB.84.075150}{Phys. Rev. B {\bf 84}, 075150 (2011).}

\bibitem{AS.2005}
F.~B. Anders and A.~Schiller,
\href{http://link.aps.org/doi/10.1103/PhysRevLett.95.196801}{Phys. Rev. Lett. {\bf 95}, 196801 (2005).}

\bibitem{AS.2006}
F.~B. Anders and A.~Schiller,
\href{http://link.aps.org/doi/10.1103/PhysRevB.74.245113}{Phys. Rev. B {\bf 74}, 245113 (2006).}

\bibitem{PSS.2010}
M.~Pletyukhov, D.~Schuricht, and H.~Schoeller,
\href{http://link.aps.org/doi/10.1103/PhysRevLett.104.106801}{Phys. Rev. Lett. {\bf 104}, 106801 (2010).}

\bibitem{WK.2010}
P.~Wang and S.~Kehrein,
\href{http://link.aps.org/doi/10.1103/PhysRevB.82.125124}{Phys. Rev. B {\bf 82}, 125124 (2010).}

\bibitem{APSSB.2011}
S.~Andergassen, M.~Pletyukhov, D.~Schuricht, H.~Schoeller, and L.~Borda,
\href{http://link.aps.org/doi/10.1103/PhysRevB.83.205103}{Phys. Rev. B {\bf 83}, 205103 (2011).}

\bibitem{KJKM.2012}
D.~M. Kennes, S.~G. Jakobs, C.~Karrasch, and V.~Meden,
\href{http://link.aps.org/doi/10.1103/PhysRevB.85.085113}{Phys. Rev. B {\bf 85}, 085113 (2012).}

\bibitem{ESGA.2012}
E.~Eidelstein, A.~Schiller, F.~Guttge, and F.~B. Anders,
\href{http://link.aps.org/doi/10.1103/PhysRevB.85.075118}{Phys. Rev. B {\bf 85}, 075118 (2012).}

\bibitem{O.2007}
X.~Oriols,
\href{http://link.aps.org/doi/10.1103/PhysRevLett.98.066803}{Phys. Rev. Lett. {\bf 98}, 066803 (2007).}

\bibitem{ASO.2009}
G.~Albareda, J.~Sune, and X.~Oriols,
\href{http://link.aps.org/doi/10.1103/PhysRevB.79.075315}{Phys. Rev. B {\bf 79}, 075315 (2009).}

\bibitem{V.2004}
G.~Vidal,
\href{http://link.aps.org/doi/10.1103/PhysRevLett.93.040502}{Phys. Rev. Lett. {\bf 93}, 040502 (2004).}

\bibitem{S.2004}
P.~Schmitteckert,
\href{http://link.aps.org/doi/10.1103/PhysRevB.70.121302}{Phys. Rev. B {\bf 70}, 121302 (2004).}

\bibitem{WF.2004}
S.~R. White and A.~E. Feiguin,
\href{http://link.aps.org/doi/10.1103/PhysRevLett.93.076401}{Phys. Rev. Lett. {\bf 93}, 076401 (2004).}

\bibitem{BSS.2010}
A.~Bransch{\"a}del, G.~Schneider, and P.~Schmitteckert,
\href{http://onlinelibrary.wiley.com/doi/10.1002/andp.201000017/abstract}{Ann. Phys. {\bf 522}, 657 (2010).}

\bibitem{S.2011}
U.~Schollw{\"o}ck,
\href{http://www.sciencedirect.com/science/article/pii/S0003491610001752}{Ann. Phys. {\bf 326}, 96 (2011).}

\bibitem{WOM.2009}
P.~Werner, T.~Oka, and A.~J. Millis,
\href{http://link.aps.org/doi/10.1103/PhysRevB.79.035320}{Phys. Rev. B {\bf 79}, 035320 (2009).}

\bibitem{WOEM.2009}
P.~Werner, T.~Oka, M.~Eckstein, and A.~J. Millis,
\href{http://link.aps.org/doi/10.1103/PhysRevB.81.035108}{Phys. Rev. B {\bf 81}, 035108 (2010).}

\bibitem{cuevassheerbook}
See for instance, J. C. Cuevas and E. Scheer, {\em Molecular Electronics: An
  Introduction to Theory and Experiments}, World Scientific, 2010.

\bibitem{jauho}
A.-P. Jauho, N.~S. Wingreen, and Y.~Meir,
\href{http://link.aps.org/doi/10.1103/PhysRevB.50.5528}{Phys. Rev. B {\bf 50}, 5528 (1994).}

\bibitem{stefanucci-rts}
G.~Stefanucci and C.-O. Almbladh,
\href{http://link.aps.org/doi/10.1103/PhysRevB.69.195318}{Phys. Rev. B {\bf 69}, 195318 (2004).}

\bibitem{perfetto}
E.~Perfetto, G.~Stefanucci, and M.~Cini,
\href{http://link.aps.org/doi/10.1103/PhysRevB.78.155301}{Phys. Rev. B {\bf 78}, 155301 (2008).}

\bibitem{svlbook}
G.~Stefanucci and R.~van Leeuwen,
\newblock {\em Nonequilibrium Many-Body Theory of Quantum systems: A Modern
  Introduction},
\newblock Cambridge Univerisity Press, 2013.

\bibitem{rikuproc}
R.~Tuovinen, R.~van Leeuwen, E.~Perfetto, and G.~Stefanucci,
\href{http://iopscience.iop.org/1742-6596/427/1/012014}{J. Phys.: Conf. Ser. {\bf 427}, 012014 (2013).}

\bibitem{CGPNG.2009}
A.~H.~C. Neto, F.~Guinea, N.~M.~R. Peres, K.~S. Novoselov, and A.~K. Geim,
\href{http://link.aps.org/doi/10.1103/RevModPhys.81.109}{Rev. Mod. Phys. {\bf 81}, 109 (2009).}

\bibitem{katsnelsonbook}
M.~I. Katsnelson,
\newblock {\em Graphene: Carbon in Two Dimensions},
\newblock Cambridge Univerisity Press, 2012.

\bibitem{onipko}
A.~Onipko,
\href{http://link.aps.org/doi/10.1103/PhysRevB.78.245412}{Phys. Rev. B {\bf 78}, 245412 (2008).}

\bibitem{1468-6996-11-5-054504}
K.~Wakabayashi, K.~Sasaki, T.~Nakanishi, and T.~Enoki,
\href{http://iopscience.iop.org/1468-6996/11/5/054504}{ Sci. Tech. Adv. Mater. {\bf 11}, 054504 (2010).}

\bibitem{XKZJZYC.2013}
H.~Xie et~al.,
\href{http://onlinelibrary.wiley.com/doi/10.1002/pssb.201349247/abstract}{Phys. Status Solidi B {\bf 250}, 2481 (2013).}

\bibitem{PSC.2010}
E.~Perfetto, G.~Stefanucci, and M.~Cini,
\href{http://link.aps.org/doi/10.1103/PhysRevB.82.035446}{ Phys. Rev. B {\bf 82}, 035446 (2010).}

\bibitem{kb}
L.~P. Kadanoff and G.~Baym,
\newblock {\em Quantum Statistical Mechanics},
\newblock Benjamin, 1962.

\bibitem{expint}
V.~Pegoraro and P.~Slusallek,
\href{http://dx.doi.org/10.1080/2151237X.2011.617177}{ Journal of Graphics, GPU and Game Tools, {\bf 15}, 183 (2011).}

\bibitem{T.2011}
I.~V. ~Tokatly,
\href{http://link.aps.org/doi/10.1103/PhysRevB.83.035127}{Phys. Rev. B {\bf 83}, 035127 (2011).}

\bibitem{AriHarju}
Y.~Hancock, A.~Uppstu, K.~Saloriutta, A.~Harju, and M.~J. Puska,
\href{http://link.aps.org/doi/10.1103/PhysRevB.81.245402}{Phys. Rev. B {\bf 81}, 245402 (2010).}

\bibitem{Schomerus}
H.~Schomerus,
\href{http://link.aps.org/doi/10.1103/PhysRevB.76.045433}{Phys. Rev. B {\bf 76}, 045433 (2007).}

\bibitem{Blanter}
Y. ~M. Blanter and I.~Martin, 
\href{http://link.aps.org/doi/10.1103/PhysRevB.76.155433}{Phys. Rev. B {\bf 76}, 155433 (2007). }

\bibitem{width}
K.~Wakabayashi, Y.~Takane, M.~Yamamoto, and M.~Sigrist,
\href{http://stacks.iop.org/1367-2630/11/i=9/a=095016}{New J. Phys. {\bf 11}, 095016 (2009).}

\bibitem{blackman}
R.~B. Blackman and J.~W. Tukey,
\newblock {\em Particular Pairs of Windows: In The Measurement of Power
  Spectra, From the Point of View of Communications Engineering (pp. 98--99)},
\newblock Dover, 1959.

\bibitem{PhysRevLett.100.206803}
X.~Wang et~al.,
\href{http://link.aps.org/doi/10.1103/PhysRevLett.100.206803}{Phys. Rev. Lett. {\bf 100}, 206803 (2008).}

\bibitem{0957-4484-22-26-265201}
M.-W. Lin et~al.,
\href{http://iopscience.iop.org/0957-4484/22/26/265201}{Nanotechnology {\bf 22}, 265201 (2011).}

\bibitem{Science.319.1229}
X.~Li, X.~Wang, L.~Zhang, S.~Lee, and H.~Dai,
\href{http://www.sciencemag.org/content/319/5867/1229.abstract}{Science {\bf 319}, 1229 (2008).}

\bibitem{doi:10.1021/nl2023756}
Y.~Lu, C.~A. Merchant, M.~Drndi{\'c}, and A.~T.~C. Johnson,
\href{http://pubs.acs.org/doi/abs/10.1021/nl2023756?journalCode=nalefd&quickLinkVolume=11&quickLinkPage=5184&selectedTab=citation&volume=11}{Nano Letters {\bf 11}, 5184 (2011).}

\bibitem{doi:10.1021/nl070133j}
Q.~Yan et~al.,
\href{http://pubs.acs.org/doi/abs/10.1021/nl070133j?journalCode=nalefd&quickLinkVolume=7&quickLinkPage=1469&selectedTab=citation&volume=7}{Nano Letters {\bf 7}, 1469 (2007).}

\bibitem{doi:10.1109/TED.2007.891872}
G.~Liang et~a.l,
\href{http://ieeexplore.ieee.org/xpls/abs_all.jsp?arnumber=4142887}{IEEE Trans. Electron Dev. {\bf 54}, 677 (2007).}

\bibitem{doi:10.1038/nnano.2008.268}
I.~Meric et~al.,
\href{http://www.nature.com/nnano/journal/v3/n11/abs/nnano.2008.268.html}{Nature Nanotech. {\bf 3}, 654 (2008).}

\bibitem{PhysRevLett.100.206802}
Z.~Li, H.~Qian, J.~Wu, B.-L. Gu, and W.~Duan,
\href{http://link.aps.org/doi/10.1103/PhysRevLett.100.206802}{Phys. Rev. Lett. {\bf 100}, 206802 (2008).}

\bibitem{animation}
{See Supplemental Material at \href{http://link.aps.org/supplemental/10.1103/PhysRevB.89.085131}{http://link.aps.org/supplemental/10.1103/PhysRevB.89.085131} for
the animations corresponding to Figs.~\ref{fig:paritya} and~\ref{fig:parityz}.}

\bibitem{petterson}
P.~Myllyperki{\"o} et~al.,
\href{http://pubs.acs.org/doi/abs/10.1021/nn1015067}{ACS Nano {\bf 4}, 6780 (2010).}

\end{thebibliography}
\end{document}